\documentclass[fleqn,usenatbib]{mnras}

\usepackage{newtxtext,newtxmath}

\usepackage[T1]{fontenc}

\DeclareRobustCommand{\VAN}[3]{#2}
\let\VANthebibliography\thebibliography
\def\thebibliography{\DeclareRobustCommand{\VAN}[3]{##3}\VANthebibliography}

\usepackage{graphicx}	
\usepackage{amsmath}	
\usepackage{subfigure}
\usepackage{multirow}
\newcommand\be{\begin{equation}}
\newcommand\e{\end{equation}}
\newcommand\ba{\begin{eqnarray}}
\newcommand\ay{\end{eqnarray}}
\newcommand\nn{\nonumber}

\newcommand\cab{C_{2, 1}^{(0)}}

\newcommand\cad{C_{2, 3}^{(0)}}

\newcommand\ccb{C_{2, 1}^{(2)}}
\newcommand\ccc{C_{2, 2}^{(2)}}
\newcommand\ccd{C_{2, 3}^{(2)}}
\newcommand\cce{C_{2, 4}^{(2)}}
\newcommand\ccf{C_{2, 5}^{(2)}}
\newcommand\pd{\partial}
\newcommand\eps{\epsilon}

\newcommand\pss{|\psi|^2}

\title[DIA Dromions in Saturn's magnetosphere]{Dust ion-acoustic dromions in Saturn's magnetosphere}

\author[K. Singh et al.]{
Kuldeep Singh,$^{1}$\thanks{E-mail: singh.kdeep07@gmail.com; kuldeep.singh@ku.ac.ae (KS).}
Michael McKerr,$^{2}$\thanks{This work was initiated during a research visit by MMK to the Department of Sciences and Engineering, Sorbonne University Abu Dhabi, UAE.}
and Ioannis Kourakis$^{3,1}$\thanks{IK was with the Department of Sciences and Engineering, Sorbonne University Abu Dhabi, UAE, during the initial phase of this research.}
\\
$^{1}$ Department  of  Mathematics,  Khalifa  University  of  Science  \&  Technology,  Abu  Dhabi,  UAE\\
$^{2}$ 14 Portna Road, Kilrea, County Derry, BT51 5SW, Northern Ireland, UK.\\
$^{3}$ Space and Planetary Science Center, Khalifa University, Abu Dhabi, UAE.\\
}

\date{Accepted XXX. Received YYY; in original form ZZZ}

\pubyear{2015}

\begin{document}
\label{firstpage}
\pagerange{\pageref{firstpage}--\pageref{lastpage}}
\maketitle

\begin{abstract}
Motivated by observations of localized electrostatic wavepackets by the Cassini -- and (earlier) by Voyager 1 and 2 -- mission(s) in Saturn's magnetosphere, we have investigated the existence conditions and the dynamical evolution of localized multi-dimensional structures in the Saturnian dusty plasma environment. To this effect, we have adopted a plasma-fluid model for dust-ion acoustic (DIA) excitations, taking into account the presence of a highly energetic (suprathermal, kappa-distributed) electron population in combination with massive dust dust particulates in the background. A multiple scales perturbation method was shown to lead to a Davey–Stewartson (DS) system of evolution equations,
if one assumes perpendicular carrier wave propagation across the magnetic field (direction). The system is then shown to possess two regimes mainly, known in the literature as DS-I and DS-II. In the former case, if certain conditions are fulfilled, exponentially localized solutions are obtained, known as dromions. The combined effects of various physical parameters such as the electron spectral index, the ambient magnetic field (strength) and the dust concentration have been examined. A numerical investigation reveals that the dromion amplitude increases with higher dust concentration, while it decreases for lower $\kappa_e$ (i.e. with an increase in the suprathermal electron population component). A stronger magnetic field results in higher amplitude but narrower dromions. Our results  provide a comprehensive framework for modeling modulated electrostatic wavepackets, in direct comparison  with experimental data in planetary environments, in Saturn's magnetosphere and elsewhere.
\end{abstract}




\section{Introduction}

Thanks to a series of observations by Voyager 1 and Voyager 2 \citep{smith81,smith82,goe83,goe89}, the ubiquitous presence of dust in Saturn's rings has by now been established as a fact.  In a wider context too, the presence of massive charged dust particulates in Space and in astrophysical plasma environments (e.g., in planetary rings, cometary tails, etc.) \citep{goe89,hor86,ver96} and also in laboratory plasmas (e.g., semiconductor chips, fusion devices, plasma devices, solar cells,  etc.) \citep{sam01,sam05,adh07} has led researchers to develop various models in order to explore a wealth of modes and instabilities that either occur only in the presence of dust, or are affected by the dust component \citep{shukla02}

A dusty plasma (DP) is generally composed of electrons and ions, as in a standard ``textbook" plasma, in addition to ionized massive dust grains bearing an electric charge (positive or negative). One of the leading DP modes, dust-ion acoustic (DIA) waves were first predicted two decades ago \citet{shukla92} and were then observed experimentally shortly thereafter \citet{bar96}. This is essentially the ion-acoustic plasma mode, where the electron thermal pressure provides the restoring force and the ion mass provides the inertia; however, the DIA wave phase speed exceeds the usual ion-acoustic speed by far, due to the ion-to-electron charge imbalance (thanks to the dust). Various investigations of the propagation characteristics of nonlinear DIA waves have been carried out in the recent years, suggesting this is still an open dynamic area of research  \citep{ik03,ik04,mam08,mam09,ali11,saini13,jen14,Laz18,mishra21}. Recently, \citet{Ksingh22} used fluid simulations to investigate the dynamics of dust-acoustic waves in the vicinity of Saturn's magnetosphere. Those simulation results were shown to be in good agreement with the theory as well as with observations of solitary waves in magnetosphere of Saturn by \citet{pic15}.

Observational evidence provided by various planetary spacecraft missions has established the ubiquitous occurrence of energetic particles in Space plasma environments, with velocity distribution presenting a  long-tailed behavior in the suprathermal component, thus diverging from the --traditionally assumed-- Maxwell-Boltzmann (thermal) distribution \citep{liu09,liva17,liva18}.  Suprathermal particles  have been reported inter alia in the Earth's magnetosphere \citep{fel75} and in its auroral region  \citep{laz08,men94}, in the magnetosheath \citep{mas06} and in Mercury's magnetospheric plasma too, as indicated by MESSENGER data \citep{Ho2016}. Vasyliunas was the first to postulate the so-called ``kappa'' (non-Maxwellian) velocity distribution as a heuristic formula intended to model the data from the OGO 1 and OGO 3 spacecraft in the Earth's magnetosphere \citep{vas68}. The kappa distribution  has subsequently been adopted  to model particle distribution in the solar wind \citep{arm83} and in planetary magnetospheres, such as Earth's, Saturn's and Jupiter's \citep{leu82}. 

The Cassini mission assembled  data from Saturn at a distance between 5.4 - 18 $R_s$, where $R_S$ is Saturn's radius ($R_S \approx 60,268$ km), thus showing that the observed electron populations can be well described by a kappa distribution \citep{sch08} with values as small as $\kappa \approx 2-6$ \citep{sch08}. 
Note that, for very large spectral index ($\kappa \rightarrow \infty$), the kappa distribution tends to the standard  form of the Maxwell-Boltzmann distribution. Observational data from the  Voyager 1 and 2 spacecraft have established that ions also obey a non-thermal distribution in Saturn's magnetosphere. Indeed, \citet{kri83} modeled the superthermal distribution  for  ions in Saturn's magnetosphere, showing that $\kappa$ lies between 6 to 8. 

The Cassini satellite equipped with  Radio and Plasma Wave Science (RPWS) revealed that charged dust in  Saturn's  E-ring effectively interacts with its magnetosphere \citep{wahlund09}. Observations of RPWS also unfolded that Enceladus (one of the largest moons of Saturn) plays a major role in the formation of the E-ring due to toroidal dust emissions with (dust particle) density of the order of $10^{-1}$ m$^{-3}$. In general, dust distribution follows  a power law $r_d^{-p}$  (with$p \sim 4-5$) \citep{kempf05,kempf06}, where $r_d$ is the radius of dust particulates. Therefore, Saturn's E-ring is predominantly populated with mm-sized dust grains. The Electron Spectrometer (ELS) instrument detected  nanometer-sized dust plume outbursts from the southern hemisphere of Enceladus \citep{kempf05}.  Cassini's RPWS wideband receiver data included a large number events consisting of  bipolar electrostatic (field) pulses within 10 $R_s$, in the period 2004-2008. This is a location  within Saturn's E ring and Enceladus's orbit in which dust density is very high \citep{pic15}. The plasma parameters corresponding to Saturn's magnetosphere are:
$n_{i, 0}=  1-20 cm^{-3}$, $n_{e0} = 2-45 cm^{-3}$, $Z_d = 10^3$, $n_d = 10^{-3}-10^{-1} cm^{-3}$ , $T_e = 1-10$ eV \citep{yar07}.

The  generalization of the fundamental theory of modulational instability in one dimension \citep{ik05} to higher dimensionality \citep{ds,Nishinari,Nishinari94,Fokas,aDuan,bDuan,Ghosh,xue04,yashika,Singh2022} is an open topic for debate among theoreticians.  Taking  transverse perturbations into account yields an anisotropic system, thus  significantly altering the  dynamical evolution of a modulated wavepacket (envelope). To model this qualititative mechanism, we have undertaken the study at hand, focusing on modulated dust-ion acoustic (DIA) wavepackets propagating in magnetized dusty plasmas in the presence of non-Maxwellian electrons in the background, in higher (2D or 3D) geometry. The evolution of modulated wavepackets in 2D/3D is typically described by the so-called Davey Stewartson system (DS) \citep{ds}, which is a multi-dimensional generalization of the nonlinear Schr\"{o}dinger equation (NLSE). 

In a series of papers by  \citet{Nishinari,Nishinari94} it was shown, by adopting a suitable  perturbation method, that the nonlinear generation of an ion acoustic 2D wavepacket in non-magnetized plasma is modeled by a DS-II system \citep{Nishinari}, whereas the similar formalism  employed  to magnetized plasma may lead to a DS-I system (depending on the relevant parameter values). We apply the classification of \citet{Fokas} who showed  that four kinds of DS systems exist. The latter system (DS-I) is known to possess localized  solutions in the form of \emph{dromions} \citep{Nishinari94}. These are 2D  hump-like structures which are exponentially localized. In contrast with 1D envelope soliton solutions, dromions do not conserve their form upon interaction with other dromions and hence can exchange energy \citep{Fokas,Fokas90}.

The study at hand is motivated by the above considerations, especially in the framework of electrostatic wave propagation in space plasmas, where such kinds of potential structures are often observed by satellites. Admittedly, little can be found in the existing literature in this challenging line of research. \citet{bDuan} examined the effect of higher order transverse perturbations by deriving the Davey–Stewartson system for wavepacket propagation in dusty plasma. \citet{Ghosh} investigated the  DS I equations for electron acoustic waves in the  Polar Cap Boundary Layer (PCBL) region, which exhibit exponentially localized dromion solutions, as discussed above. \citet{xue04} modeled the  modulation of DIA waves in unmagnetized  plasma  and derived a three-dimensional Davey–Stewartson (3D DS) equation. It may be emphasized that the modulation properties of DIA wavepackets in 3D dusty plasmas are unlike the 1D case. \citet{yashika} examined the dynamical properties of dust acoustic dromions in relation with observations by the POLAR satellite in the presence of superthermal electrons/ions in the polar cap boundary layer region of the Earth's atmosphere. Recently, \citet{Singh2022} investigated the stability of dust-ion acoustic 2D modulated wavepackets in Saturn's magnetosphere, taking into account the presence of charged dust and suprathermal electrons; rather counterintuitively, that plasma model did not lead to the DS-I regime, thus ruling out the existence of dromions (at least, within the cold ion fluid model adopted therein; including thermal ion pressure in the model may be expected to alter this picture).

In this article, we shall rely on a multiple scale perturbation technique to obtain a  Davey–Stewartson (DS) system of equations, in order to investigate the occurrence of dust-ion-acoustic dromions in magnetized dusty plasmas with suprathermal  electrons.   Our results can be compared with existing experimental/observational data in space (and may hopefully motivate new ones), in particular in Saturn's magnetosphere \citep{goe89}.

\section{Fluid model}

Let us consider a magnetized plasma comprising of inertial ions (mass $m_i$, charge $q_i = + Z_i e$), non-Maxwellian electrons (mass $m_e$, charge $-e$) and immobile charged  dust particles (mass $M_d \gg m_i$, charge $Q_d = s_d Z_{d0} e$,  where we have used $s_d = \pm 1$ to denote the dust charge sign, i.e. $s_{d}=+1$ for positively charged dust and $s_d=-1$ for negatively charged dust. Note that, although we have $s_d$ arbitrary here, for generality, in the study that follows, we have chosen $s_d=-1$, i.e. restricting our work to the case of negative dust. The magnetic field is assumed to be oriented along the z-axis (i.e., $\vec B=B_0 \hat{z}$.

The fluid model equations, describing the plasma state in terms of the ion fluid density $N_{i}$ and speed  $V_{i}$ as functions of time $t'$ and space $\vec{r'} = \{ x', y'\}$, include the
continuity equation: \be \frac{\pd N_{i}}{\pd t'}  + \nabla'\cdot(N_{i}\vec V_{i})=0 \, ,\label{1}\e
the momentum  equation:\be \frac{\pd\vec V_{i}}{\pd t'}+\vec V_{i}\cdot \nabla' \vec V_{i}=-\frac{Z_{i}e}{m_i}\nabla' \Phi+\frac{Z_{i}e}{m_i}(\vec V_i \times \vec B) \, ,\label{2}\e
where the electrostatic (ES) potential $\Phi$ is defined in the right-hand side (rhs) and Poisson's equation: \be \nabla'^2\Phi=- \frac{e}{\epsilon_0}  (Z_{i}N_{i}-Z_{e}N_{e}+s_{d}Z_{d0}N_{d0}) \, , \label{3}\e where $\epsilon_0$ is the permittivity of vacuum.
The electron density is given upon integrating the kappa velocity distribution by the expression \citep{hellberg09}
\be N_{e}=n_{e0}\left(1-\frac{e \Phi}{k_B T_e(\kappa_{e}-\frac{3}{2})}\right)^{-\kappa_{e}+\frac{1}{2}}\label{4} \, .\e
 The charge neutrality condition at equilibrium imposes
\be n_{e0} = Z_i n_{i, 0} + s_{d}Z_{d0} n_{d0} \, , \label{QN} \e
where $n_{\jmath0}$ for ($\jmath=i, e, d$) denote(s) the unperturbed number density for the electrons, for the ions and for the dust, respectively. Given the large mass of the dust particulates, the dust component is assumed to be stationary i.e. its density is assumed to be fixed, i.e. $n_{d} = n_{d0}  =$ constant. It is also assumed that the dust charge is constant, for simplicity (i.e. dust charging is neglected).

To facilitate further analysis, Eqs. (\ref{1})-(\ref{4}) may be re-scaled by introducing the following dimensionless variables: the number density $n_j = {N_{j}}/{n_{i, 0}}$ (for $j = i, e, d)$; velocity $v_{i} = {V_{i}}/{C_{i}}$ (i.e., $C_{i}=(Z_{i}k_{B}T_{e}/m_{i})^{1/2}$); the electrostatic potential $\phi =  \frac{e\Phi}{k_B T_e}$; time and space are respectively rescaled as $t = t' \omega_{pi}$ (where $\omega_{pi}=[e^{2}Z_{i}^{2}n_{i, 0}/(\epsilon_0 m_{i})]^{1/2}$) and $\{ x, y \} = \{x', y' \}/{\lambda_{D, i}}$, hence the space derivative operator (divergence vector) is re-scaled as $\nabla = {\lambda_{D, i}}^{-1} {\nabla'} $ (where $\lambda_{D, i}=[\epsilon_0 k_{B}T_{e}/(e^{2}Z_{i}n_{i, 0})]^{1/2}$ ).
Finally, we have defined the ratio $\Omega=\frac{\omega_{ci}}{\omega_{pi}}$, where $\omega_{ci}=\frac{Z_i e B_0}{m_i}$ is the ion cyclotron frequency. The charge neutrality requirement (at equilibrium) imposes the constraint:
\[\delta_e =\frac{n_{e0}}{Z_{i}n_{i, 0}}= 1+ s_d \delta_d \, , \]
where $\delta_{d}=\frac{Z_{d0}n_{d0}}{Z_{i}n_{i, 0}}$. (From now on, we set $s_d = -1$, i.e. for negative dust.) Note the effect of the dust on the electron-to-ion density ratio, that deviates from unity and may actually acquire significant values, thus affecting the wave's phase speed and overall characteristics --  as well known to be a unique feature of the dust-ion acoustic mode \citep{shukla02}.
Note that the classical, textbook ion-acoustic (in \emph{e-i}, i.e. dust-free plasma) is recovered for $\delta_d=0$, while the unmagnetized case can  be recovered for $\vec B= \vec 0$ \citep{Singh2022}.

In the following, all quantities will be dimensionless, unless otherwise stated. Applying the above normalization, the fluid model Eqs. (\ref{1})-(\ref{4}) become
\begin{eqnarray} \frac{\pd n_{i}}{\pd t}  +\nabla\cdot(n_{i}\vec v_{i})=0 \, ,\label{5} \\
 \frac{\pd\vec v_{i}}{\pd t}+\vec v_{i}\cdot \nabla \vec v_{i}=-\nabla\phi + \Omega \vec v_i \times \hat{z} \, , \label{6} \\
 \nabla^2\phi=n_{e} - n_{i} - s_{d}\delta_d \, ,\label{7} \end{eqnarray}
where the normalized expression for the electron density is
\be n_{e}=\delta_e\left(1-\frac{\phi}{\kappa_{e}-\frac{3}{2}}\right)^{-\kappa_{e}+\frac{1}{2}}\approx \delta_{e}+c_1\phi +c_2\phi^2+c_3\phi^3.\label{4a}\e
Here, $c_1=\delta_e\frac{\kappa_e-\frac{1}{2}}{(\kappa_e-\frac{3}{2})}$, \ $c_2=\delta_e\frac{\kappa_{e}^{2}-\frac{1}{4}}{2(\kappa_{e}-\frac{3}{2})^{2}}$ and  $c_3=\delta_e\frac{(\kappa_{e}^{2}-\frac{1}{4})(\kappa_{e}+\frac{3}{2})}{6(\kappa_{e}-\frac{3}{2})^{3}}$ are  (real) coefficients resulting from a McLaurin series expansion near equilibrium, i.e. for small $\phi$.
Note that the coefficients $c_{1,2,3}$ remain positive for all values of $\kappa_e$ and $\delta_e$ (or $\delta_d$).
Expanding the RHS near equilibrium, Poisson's Eq. (\ref{7}) thus becomes
\be \nabla^2\phi \simeq (1-n_{i})+c_1\phi +c_2\phi^2+c_3\phi^3 \, , \label{7a}\e
where all coefficient were defined above. The quasi-neutrality condition (\ref{QN}) (valid at equilibrium) was used to simplify the latter equation, which is readily seen to hold at equilibrium (upon setting $n_i = 1$ and $\phi = 0$).

\section{Perturbative analysis}

We proceed by expanding the state variables around equilibrium as
\ba \phi&=&\eps\phi_1+\eps^2\phi_2+\eps^3\phi_3+...\nn\\
 n_i&=&1+\eps n_1+\eps^2n_2+\eps^3 n_3+...\nn \\
 v_{i}&=& \eps v_{1}+\eps^2 v_{2}+\eps^3 v_{3}+... \ay
and by introducing multiple evolution scales considered for the independent (time, space) variables as $ T_j =\eps^j t$ and   $\vec{X_j}=\eps^j \vec{x}$ where $j = 0, 1, 2, 3, ...$.
At every order $\eps^j$, the state variables are expanded as
\ba \phi_j&=& \sum_{l=-j}^j\phi_{j}^{(l)}e^{i l (\vec k \vec X_0 - \omega T_0)}  \nn \\
 n_j&=& \sum_{l=-j}^jn_{j}^{(l)}e^{i l (\vec k \vec X_0 - \omega T_0)} \nn \\
 v_{j}&=& \sum_{l=-j}^jv_{j}^{(l)}e^{i l (\vec k \vec X_0 - \omega T_0)} \, , \ay
where the phase obviously depends on the zeroth-order (fast) variables, while the harmonic amplitudes are assumed to depend only on the slower scales (for $l = 1, 2, ...$). Note that the wavenumber is $\vec k=k_1\hat{x}+k_2\hat{y}+k_3\hat{z}$ and the fluid velocity (vector) is expressed as $v_{i} = v_x \hat{x}+v_y\hat{y}+v_z\hat{z}$.

\subsection{Linear analysis}

In order $\eps^1$, we obtain the following evolution equations:
\ba -\omega n_1+k_1v_{x1}+k_2v_{y1}+k_3v_{z1}&=&0\nn \\
 -\omega v_{x1}+k_1\phi_1 +i \Omega v_{y1}&=&0\nn \\
 -\omega v_{y1}+k_2\phi_1-i \Omega v_{x1}&=&0\nn \\
 -\omega v_{z1}+k_3\phi_1&=&0\nn \\
n_1&=&(c_1+k^2)\phi_1\ay
In order for a non-trivial solution to exist, a compatibility condition must be satisfied.
One thus obtains a bi-quadratic equation in terms of $\omega$ as
\be \omega^4-\omega^2\left(\Omega^2+\frac{k^2}{c_1+k^2}\right)+\frac{\Omega^2k_3^2}{c_1+k^2}=0 \, , \e
where $k^2 = k_1^2 + k_2^2 + k_3^2$ is the norm of the wavenumber (vector).
From the above equation, we obtain the dispersion relation
\be \omega_\pm^2=  \frac{1}{2}\left[\Omega^2+\frac{k^2}{c_1+k^2}\pm \sqrt{\left(\Omega^2+\frac{k^2}{c_1+k^2}\right)^2-\frac{4\Omega^2k_3^2}{c_1+k^2}}\right]\label{dispersion}\e

As one might expect, the dispersion relation is symmetric in the first two arguments $\omega(k_1, k_2, k_3)=\omega(k_2, k_1, k_3)$. To simplify the algebra further, we may define the $x-$ and $z-$ axes such that $\vec k=(k_x, 0, k_z) = (k_\perp, 0, k_\parallel)$, in terms of the wavenumber components in the directions perpendicular and parallel to the magnetic field (strength), via $\Omega$. The dispersion relation thus takes the form
\be \omega_\pm^2=  \frac{1}{2}\left[\Omega^2+\frac{k^2}{c_1+k^2}\pm \sqrt{\left(\Omega^2+\frac{k^2}{c_1+k^2}\right)^2-\frac{4\Omega^2k_\parallel^2}{c_1+k^2}}\right] \, , \label{dispersion}\e
where $k^2 = k_\perp^2 + k_\parallel^2$. This relation essentially generalizes the dispersion relation of oblique magnetosonic (magnetoacoustic) waves \citep{son03} (see pp. 107-109 therein) in the presence of non-Maxwellian electrons. (Note, for the sake of comparison with earlier works, that $c_1= 1$ in the thermal/Maxwellian electron-ion i.e. dust-free plasma limit). \color{black}
Note that an identical 3D dispersion relation was derived and analyzed earlier in \citet{williams13}; see eq. (7) therein.

Notice that the angular frequency is a function of $k_\perp$ and $k_\parallel$, essentially, that depends parametrically on the value of $\kappa_e$ (via $c_1$) and on the magnetic field. In the presence of the magnetic field, two dispersion curves, say $\omega_-$ and $\omega_+$, are obtained; note the two signs in (\ref{dispersion}) above.  The lower sign corresponds to an acoustic mode, while the upper sign leads to an optic-like mode, with a cutoff frequency at the cyclotron frequency, as will be discussed below.

It may be appropriate to consider certain known limits at this point.
First of all, if one considers the  unmagnetized case, i.e. $\Omega=0$, eq. (\ref{dispersion}) takes the familiar dust ion-acoustic (DIA) dispersion relation
\be \omega^2 = \frac{k^2}{c_1+k^2} \label{dispersion1} \, , \e
where $k = \sqrt{k_\perp^2 + k_\parallel^2}$, as expected. Note that this relation is isotropic, i.e. spherically-symmetric in space. This is true only in the absence of magnetic field, i.e. for $\Omega=0$.

For parallel propagation ($k_\perp = 0$, hence $k_\parallel = k$), the above equation (\ref{dispersion}) leads to either
\be \omega^2=  \Omega^2; \qquad {\rm or} \qquad  \frac{k^2}{c_1+k^2} \, . \label{dis3}\e
The former represents localized (non-propagating) Larmor oscillations, while the latter is the DIA mode above, as expected (recall that the Larmor force $\sim \mathbf{v} \times \mathbf{B}$ bears no component in the direction of the magnetic field).

For purely perpendicular propagation, equation (\ref{dispersion}) reduces to
\be \omega^2 = \Omega^2+\frac{k^2}{c_1+k^2} \label{dispersion1} \, \e
where $k =  k_\perp$ (and $k_\parallel = 0$) in this case.

\begin{figure*}
\centering
\includegraphics[width=6in]{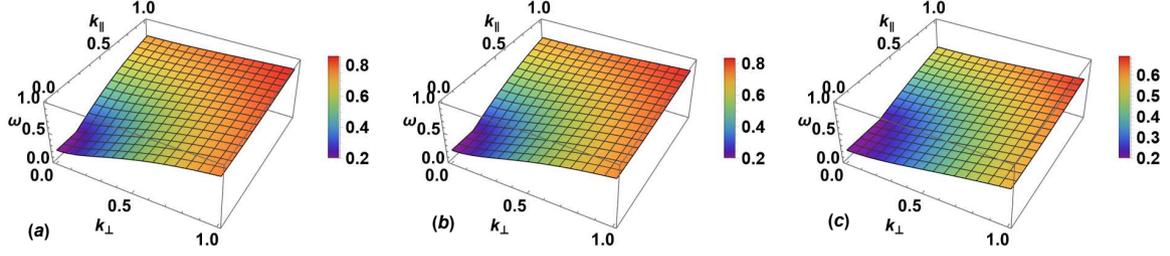}
\caption{Plot of $\omega_+$ -- obtained from (\ref{dispersion}) -- versus $k_{\perp}$ and $k_{\parallel}$, for $\delta_d=\Omega=0.2$ (fixed) and for different values of (a) $\kappa_e = 100$ (quasi-Maxwellian), (b) $\kappa_e=6$, (c) $\kappa_e=2$ (strong deviation from the Maxwellian limit).}\label{fex1}
\end{figure*}

\begin{figure*}
\centering
\includegraphics[width=6in]{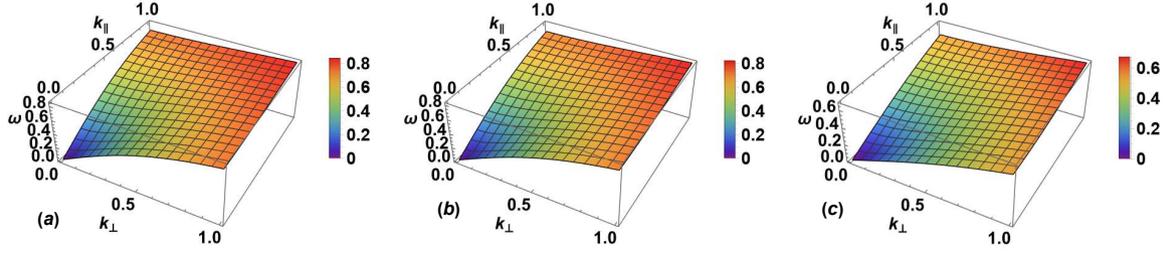}
\caption{Plot of $\omega_+$ -- obtained from (\ref{dispersion}) -- versus $k_{\perp}$ and $k_{\parallel}$, for $\Omega=0$ (unmagnetized plasma), $\delta_d=0.2$, for different values of (a) $\kappa_e=100$ (quasi-Maxwellian), (b) $\kappa_e=6$, (c) $\kappa_e=2$ (strong deviation from the Maxwellian limit).}\label{fex2}
\end{figure*}

Figure \ref{fex1} depicts the variation of the upper mode in the  $(k_{\perp}, k_{\parallel})$ plane, for the magnetized case. The analogous dispersion curve for  unmagnetized plasma is shown in  Fig \ref{fex2}. An indicative value of $\delta_d=0.2$ is taken in all plots.

The leading- (i.e. first-) order first-harmonic amplitudes can be conveniently expressed in terms of the electrostatic potential disturbance (amplitude) as:
\ba \phi_1^{(1)}&=& \psi\nn\\
n_1^{(1)}&=&(c_1+k^2)\psi=\mu_n\psi\nn\\
v_{x1}^{(1)}&=&\frac{\omega k_x}{(\omega^2-\Omega^2)}\psi=\mu_x\psi\nn\\
v_{y1}^{(1)}&=& -\frac{i\Omega}{\omega}v_{x1}^{(1)}=\mu_y\psi\nn\\
v_{z1}^{(1)}&=&\frac{k_z}{\omega}\psi=\mu_z\psi \, .\ay

\subsection{Nonlinear analysis in 2$^{nd}$-order: group velocity and harmonic generation}

In order $\eps^2$, eliminating the secular terms yields a condition in the form:
\be \frac{\pd \psi}{\pd T_1}+ v_{g,x}\frac{\partial \psi}{\partial X_1}+v_{g,z}\frac{\partial \psi}{\partial Z_1}=0\label{kramer} \, , \e
where the group velocity components are
\be  v_{g,x}=\frac{\partial \omega}{\partial k_x}=\frac{\frac{\omega^2 c_1}{(c_1+k^2)^2}+\frac{\Omega^2k_z^2}{(c_1+k^2)^2}}{\omega\left(2\omega^2-\left(\Omega^2+\frac{k^2}{c_1+k^2}\right)\right)}k_{x}\, \label{vg1}\e
and
\be  v_{g,z}=\frac{\partial \omega}{\partial k_z}=\frac{\frac{\omega^2 c_1}{(c_1+k^2)^2}-\frac{\Omega^2(c_1+k_x^2)}{(c_1+k^2)^2}}{\omega\left(2\omega^2-\left(\Omega^2+\frac{k^2}{c_1+k^2}\right)\right)}k_{z}\, \label{vg2}\e
The above algebraic constraint dictates that the amplitude(s) of all  harmonics, at this order, will rely (only) on the moving coordinates $X_1- v_{g,x} T_1$ and $Z_1-v_{g,z} T_1$;  physically speaking, the first harmonic envelope (amplitude) will move at the group velocity in this order, viz. 
$\psi = \psi(X_1- v_{g, x} T_1; Z_1 - v_{g, z} T_1; X_2, Z_2, T_2)$ for the electrostatic potential (with analogous expressions for the first-harmonic amplitudes of all other state variables) \citep{Singh2022}.
This type of variable transformation has already been employed in a numerous articles on nonlinear plasma wave dynamics -- see e.g. \citet{InfeldRowlands} or \citet{ik05} in the context of space physics and also in nonlinear optics; see e.g. \citet{NewellBook}.

After solving the equations obtained to 2$^{nd}$-order, a set of  zeroth, first and second harmonics expressions are derived. It is easy to express all quantities (like  density and fluid speed)  in terms of the electrostatic potential, for each harmonic. One may take $\phi_2^{(1)}=0$ without any loss of generality, hence the first harmonic amplitudes are obtained as:
\ba n_2^{(1)}&=&-2i\vec k\cdot\nabla_1\psi\, , \nn\\
v_{x2}^{(1)}&=&-\frac{\omega}{\omega^2-\Omega^2}\left(i\mu_x\left(1+\frac{\Omega^2}{\omega^2}\right)\frac{\pd\psi}{\pd T_1}+i\frac{\pd\psi}{\pd X_1}+ 2ik_xk_z\frac{\pd\psi}{\pd Z_1}\right)\, , \nn\\
v_{y2}^{(1)}&=&-\frac{i\Omega}{\omega}v_{x2}^{(1)}-\frac{i\mu_y}{\omega}\frac{\pd\psi}{\pd T_1}-\frac{i}{\omega}\frac{\pd\psi}{\pd Y_1}\, , \nn\\
v_{z2}&=& -\frac{i\mu_z}{\omega}\frac{\pd\psi}{\pd T_1}-\frac{i}{\omega}\frac{\pd\psi}{\pd Z_1} \, . \nn\\ \ay

From the  2$^{nd}$ order 2$^{nd}$ harmonics, we obtain the respective second harmonic amplitudes as:
\ba\phi_2^{(2)}&=&\frac{-\left[\omega^2c_2(4\omega^2-\Omega^2)-\chi(c_1+k^2)^2\right]\psi^2}{3\left(4\omega^4k^2+\Omega^2(c_1\omega^2-k_z^2)\right)}=\ccd\psi^2 \nn \\
 n_2^{(2)}&=&\left((c_1+4k^2)\ccd+c_2\right)\psi^2=\ccb\psi^2\nn\\
 v_{z2}^{(2)}&=&\frac{k_z}{\omega}\left(\ccd+\frac{c_1+k^2}{2}\right)\psi^2=\ccc\psi^2\nn\\
 v_{x2}^{(2)}&=&\frac{4\omega k_x}{4\omega^2-\Omega^2}\left(\ccd+\frac{\omega^2(c_1+k^2)}{2(\omega^2-\Omega^2)}\right)=\cce\psi^2\nn\\
v_{y2}^{(2)}&=&-\left(\frac{i\Omega}{2\omega}v_{x2}-\frac{i\Omega k_x}{2}\frac{(c_1+k^2)}{2(\omega^2-\Omega^2)}\right)\psi^2=\ccf\psi^2\label{ccd}\ay
where $\chi= 6\omega^2(\omega^2-\Omega^2)+\frac{3\omega^2}{2}\left(1+2\Omega^2+\frac{k_x^2}{\omega^2-\Omega^2}\right)\left(\frac{1}{c_1+k^2}\right)$. The zeroth harmonic amplitudes (to second order) are not conclusively determined this order, so one needs to resort to  the third order equations ($\eps^3$) to find their analytical expression.

The expanded fluid equations at zeroth order can then be solved in terms of $\pss$ and $\Phi=\int dX_1\frac{\pd v_{z2}}{\pd Z_1}$ to find:
\ba\phi_2&=&\left(2\frac{\mu_n\mu_x}{c_1v_{g,x}}-\frac{2c_2}{c_1}\right)\pss+\frac{1}{c_1v_{g,x}}\Phi=\cad\pss+\gamma_\phi \Phi\nn\\
n_2^{(0)}&=&\frac{2\mu_n\mu_x}{v_{g,x}}\pss+\frac{1}{v_{g,x}}Y=\cab\pss+\gamma_n \Phi\label{cad}\ay
The integration constants in the above expressions are set to zero. If this was not the case, then a term proportional to $\psi$ would appear in the first equation of the DS system, which could then be removed by a phase shift on $\psi$.

Note the appearance in the latter expressions of a new quantity, $\Phi$, that was absent (i.e. vanishes) in the one-dimensional modulation case.

\begin{figure*}
\centering
\includegraphics[width=6in]{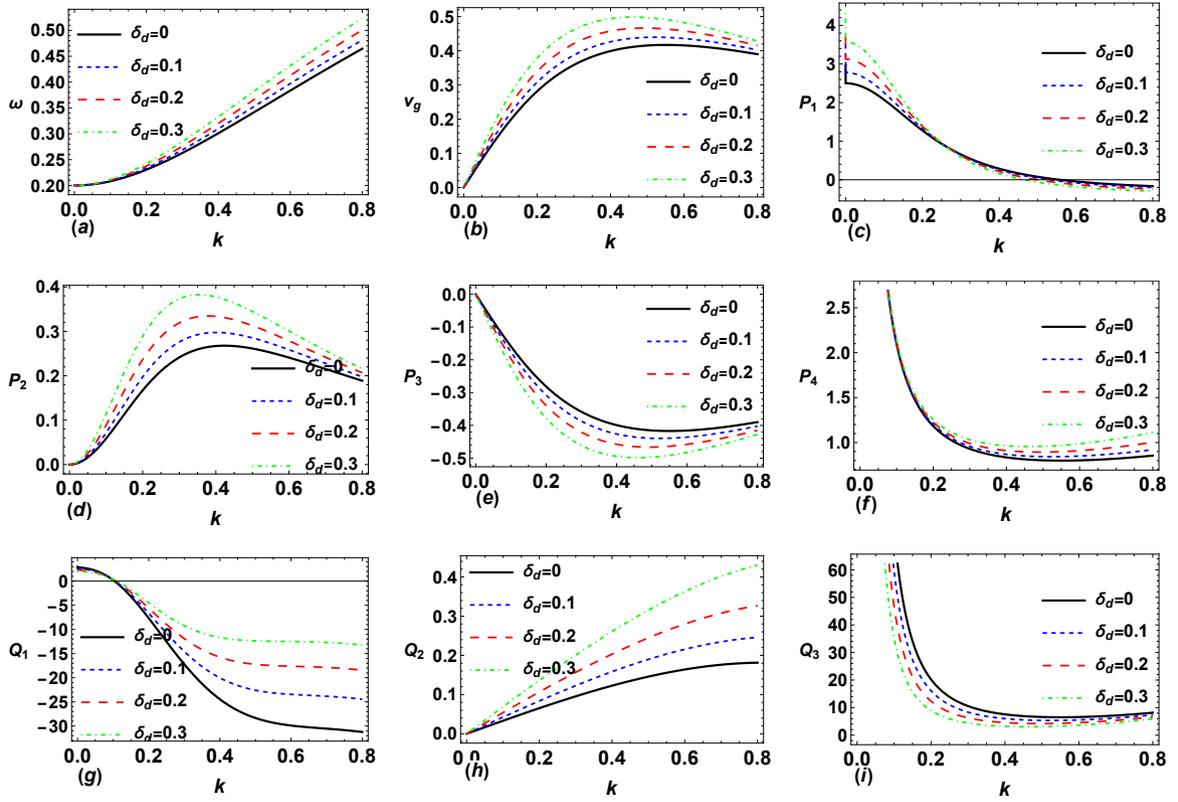}
\caption{Plot of (a) $\omega_+$ vs. $k$ (b) $v_g$ vs. $k$ (c) $P_1$  vs. $k$ (d) $P_2$  vs. $k$ (e) $P_3$  vs. $k$ (f) $P_4$  vs. $k$ (g) $Q_1$  vs. $k$ (h) $Q_2$  vs. $k$ (i) $Q_3$  vs. $k$  for different values of $\delta_d$ and for fixed values of $\kappa_e=2$ and $\Omega=0.2$ . Note that the upper mode $\omega_+$ (only) has been taken into account in all plots.}\label{f1}
\end{figure*}

\begin{figure*}
\centering
\includegraphics[width=6in]{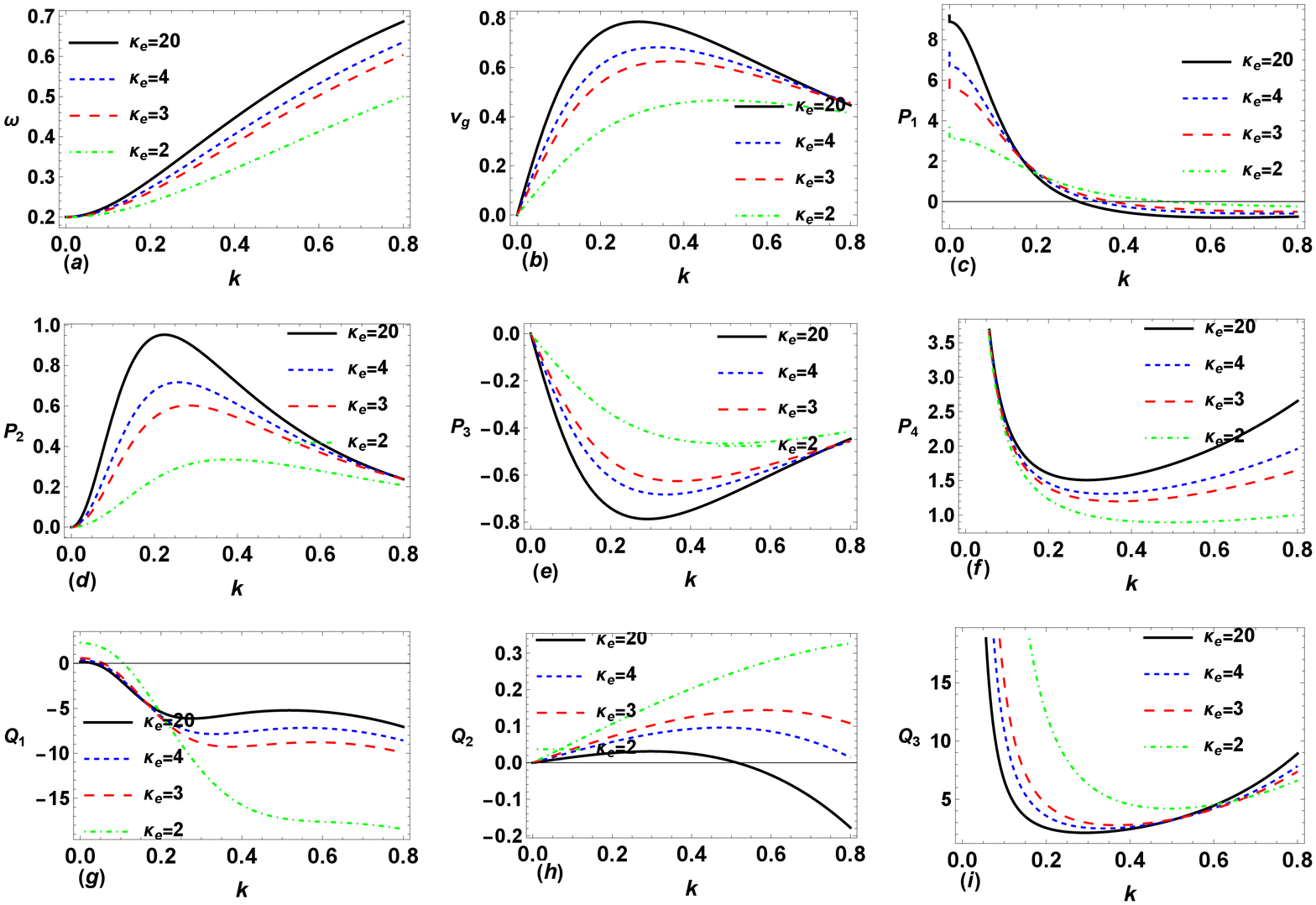}
\caption{Plot of (a) $\omega_+$ vs. $k$ (b) $v_g$ vs. $k$ (c) $P_1$  vs. $k$ (d) $P_2$  vs. $k$ (e) $P_3$  vs. $k$ (f) $P_4$  vs. $k$ (g) $Q_1$  vs. $k$ (h) $Q_2$  vs. $k$ (i) $Q_3$  vs. $k$  for different values of $\kappa_e$ and for fixed values of $\delta_d=0.2$ and $\Omega=0.2$ . Note that the upper mode $\omega_+$ (only) has been taken into account in all plots.}\label{f2}
\end{figure*}

\begin{figure*}
\centering
\includegraphics[width=6in]{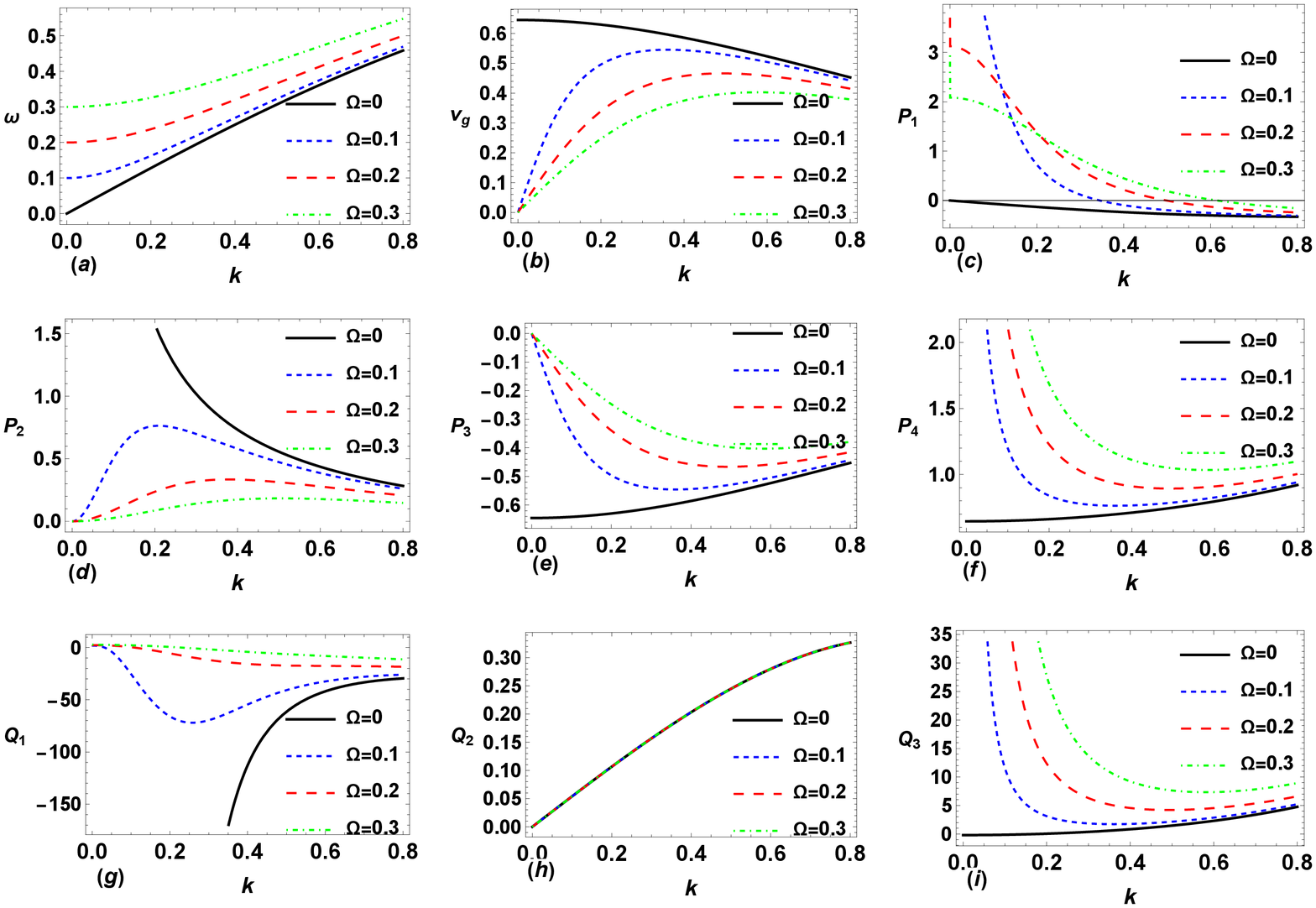}
\caption{Plot of (a) $\omega_+$ vs. $k$ (b) $v_g$ vs. $k$ (c) $P_1$  vs. $k$ (d) $P_2$  vs. $k$ (e) $P_3$  vs. $k$ (f) $P_4$  vs. $k$ (g) $Q_1$  vs. $k$ (h) $Q_2$  vs. $k$ (i) $Q_3$  vs. $k$  for different values of $\Omega$ and for fixed values of $\delta_d=0.2$ and $\kappa_e=2$ . Note that the upper mode $\omega_+$ (only) has been taken into account in all plots.}\label{f3}
\end{figure*}

\begin{figure*}
\centering
\subfigure[]{\includegraphics[width=2.5in]{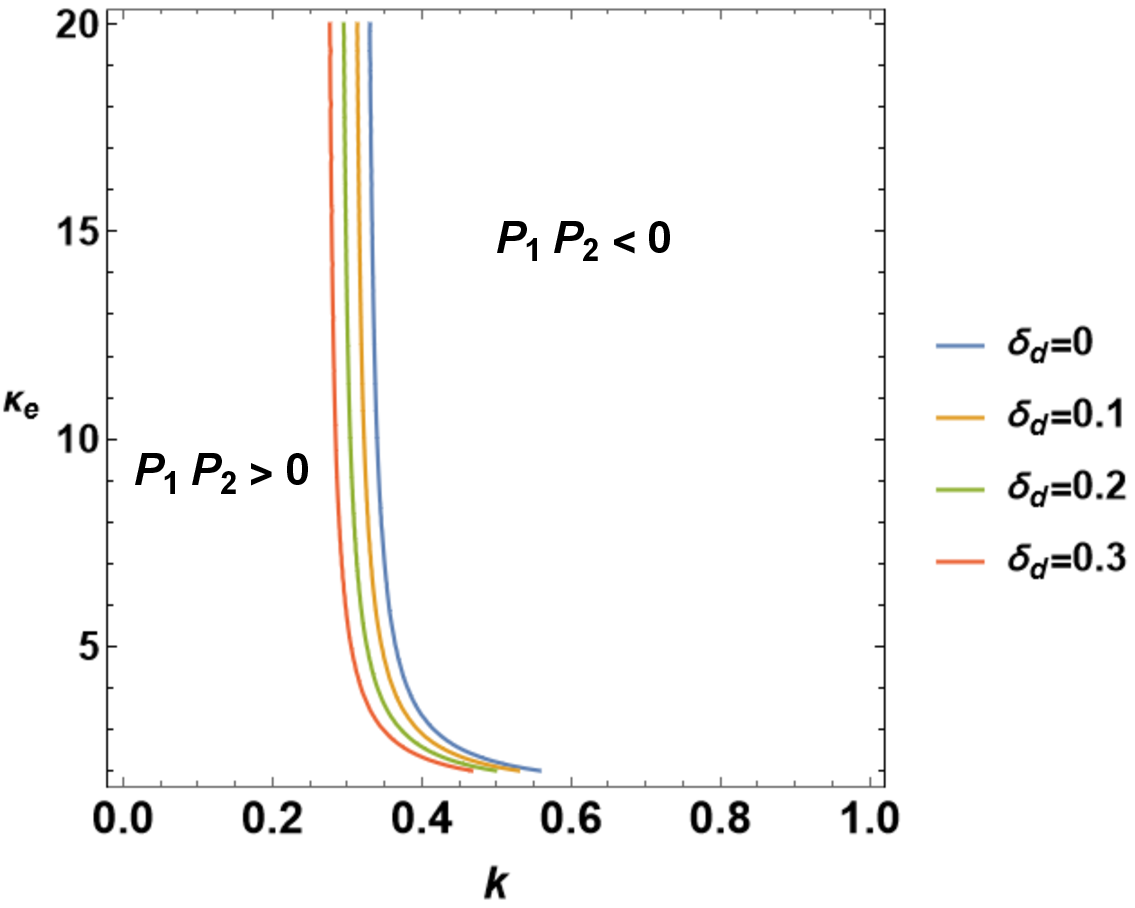}}
\subfigure[]{\includegraphics[width=2.5in]{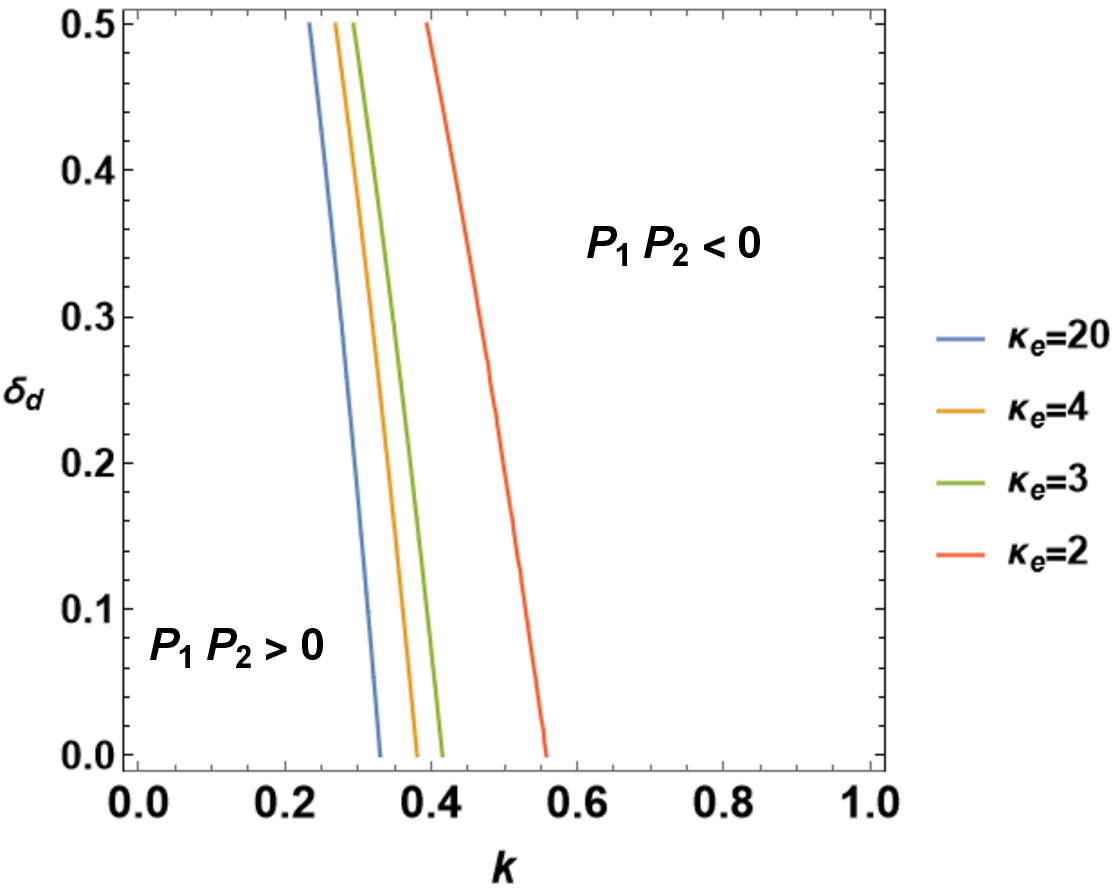}}
\subfigure[]{\includegraphics[width=2.5in]{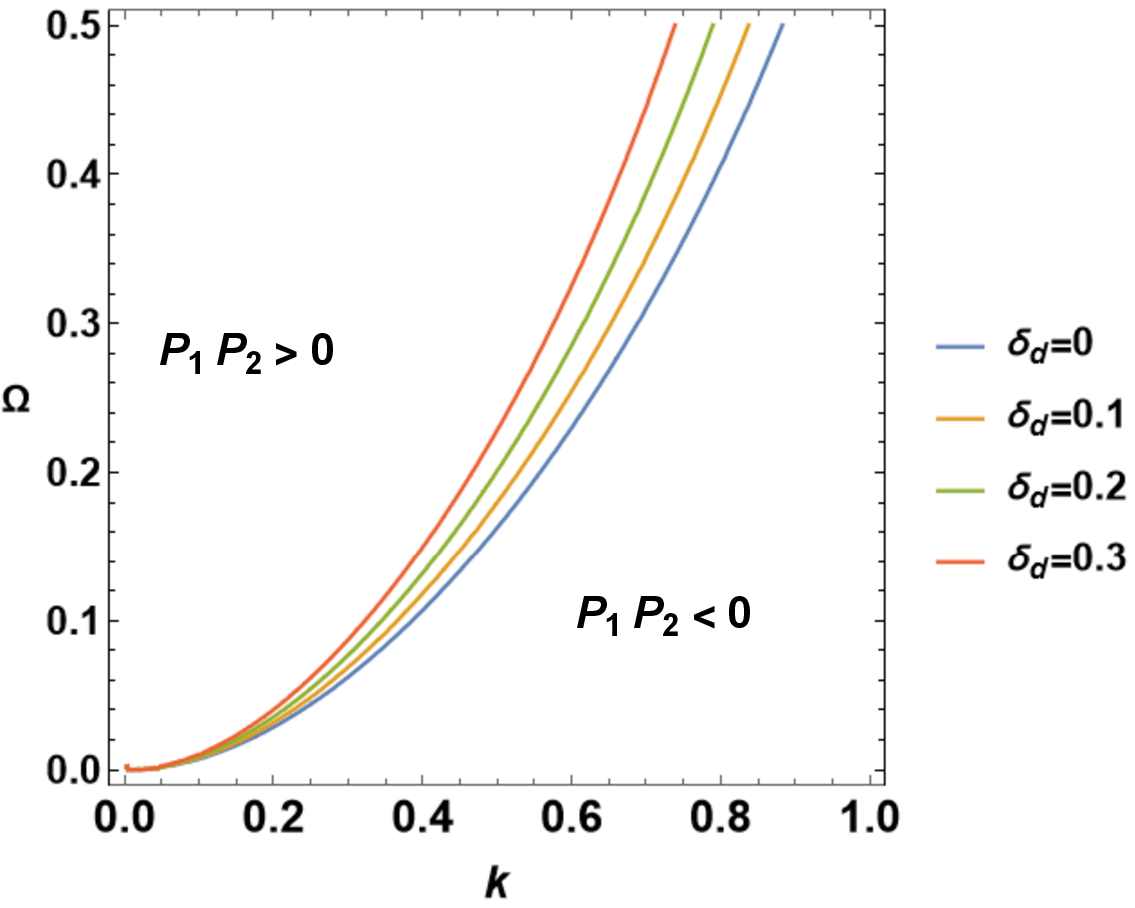}}
\caption{ContourPlot of $P_1P_2$ (a) in $k-\kappa_e$ plane for different $\delta_d$ and for a fixed value of $\Omega=0.2$; (b) in the $k-\delta_d$ plane, for different $\kappa_e$, for fixed values of $\Omega=0.2$; (c) in the  $k-\Omega$ plane, for different values of $\delta_d$ and for a fixed value of $\kappa_e=2$.}\label{f4}
\end{figure*}
\begin{figure*}
\centering
\subfigure[]{\includegraphics[width=2.5in]{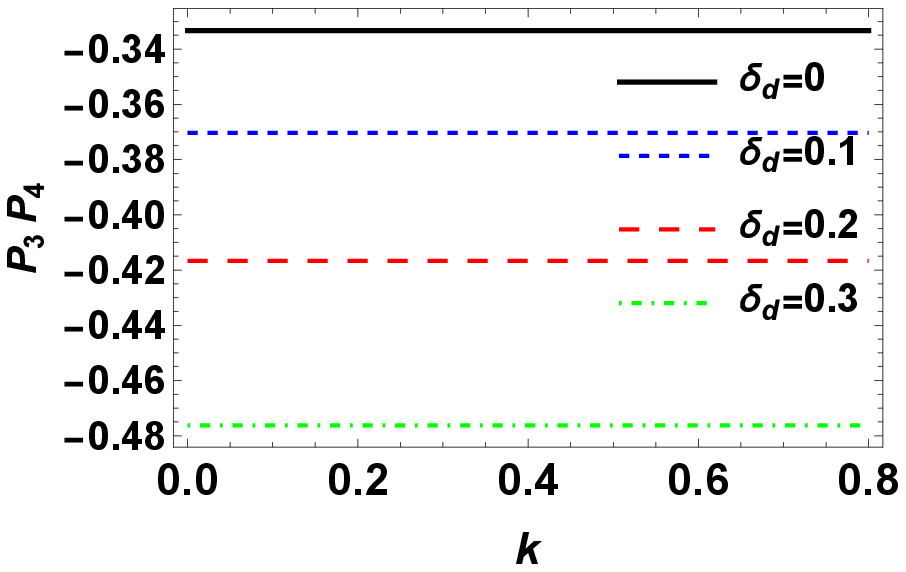}}
\subfigure[]{\includegraphics[width=2.5in]{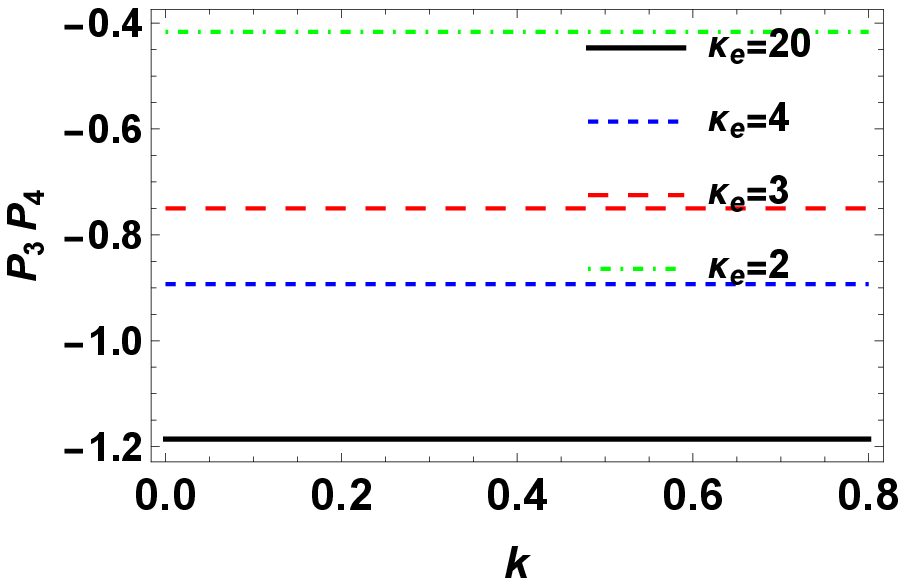}}
\subfigure[]{\includegraphics[width=2.5in]{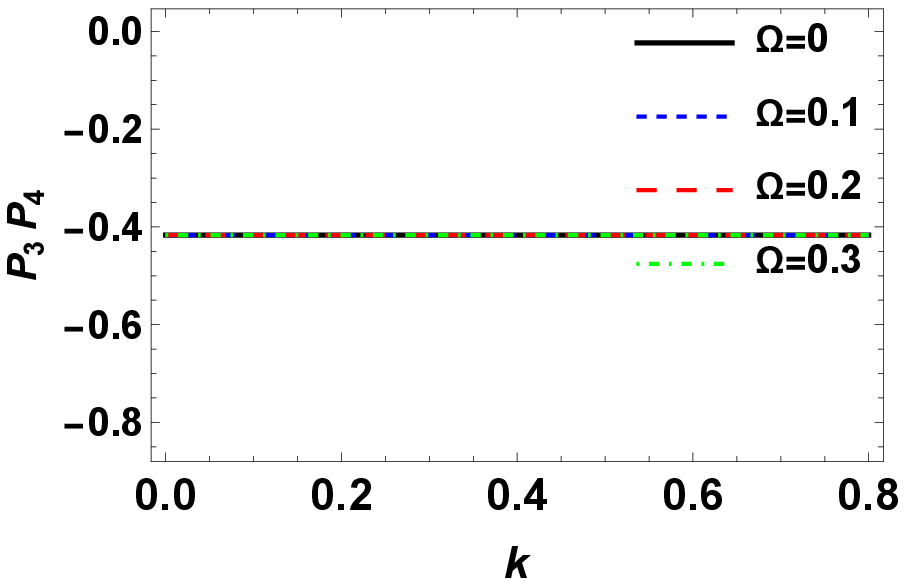}}
\caption{Plot of $P_3P_4$ vs. $k$ for different values of (a) $\delta_d$,  for fixed values of $\kappa_e=2$, $\Omega=0.2$; (b) $\kappa_e$, for fixed values of  $\Omega=0.2$, $\delta_d=0.2$; (c) $\Omega$, for fixed values of $\kappa_e=2$, $\delta_d=0.2$.}\label{f5}
\end{figure*}

\begin{figure*}
\centering
\subfigure[]{\includegraphics[width=2.5in]{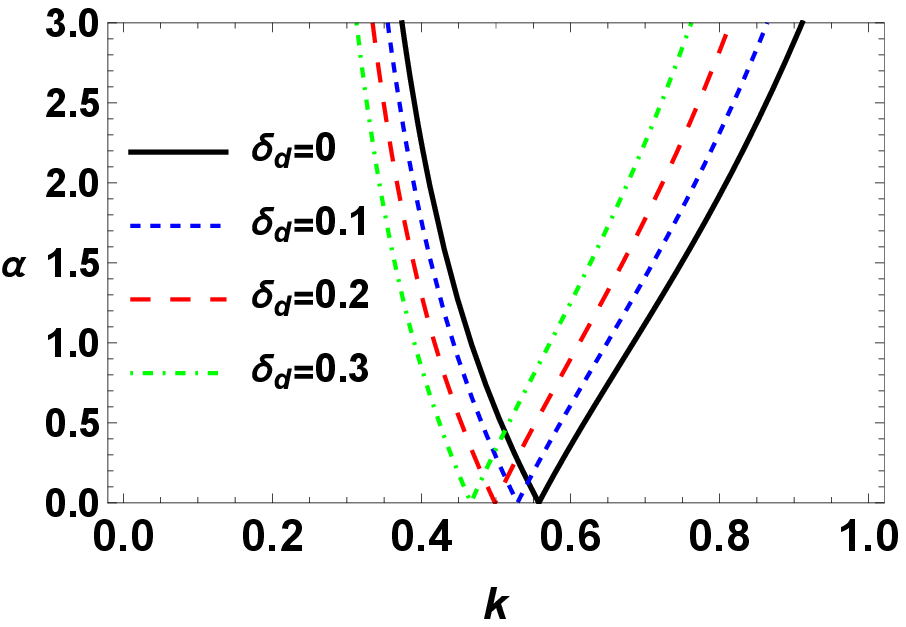}}
\subfigure[]{\includegraphics[width=2.5in]{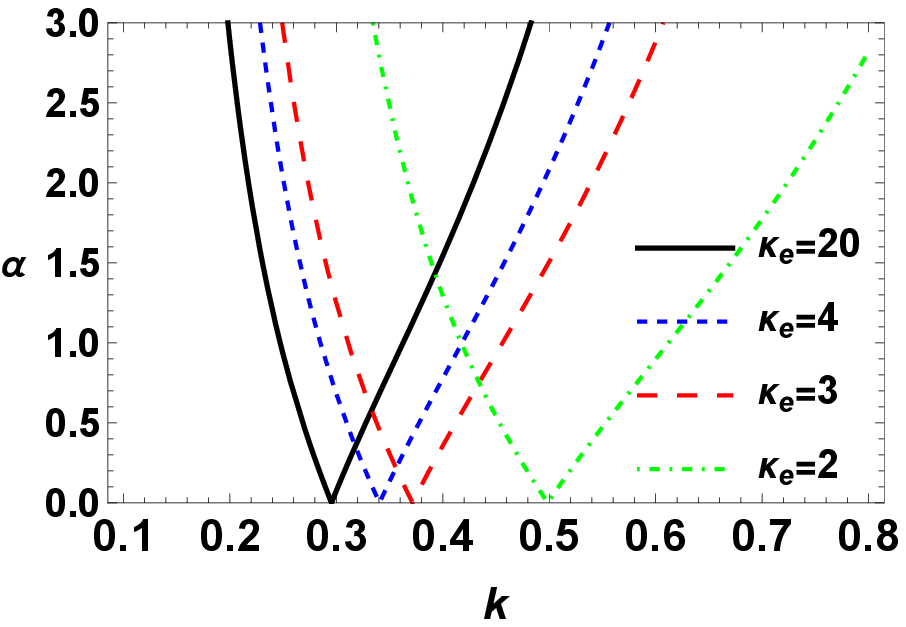}}
\subfigure[]{\includegraphics[width=2.5in]{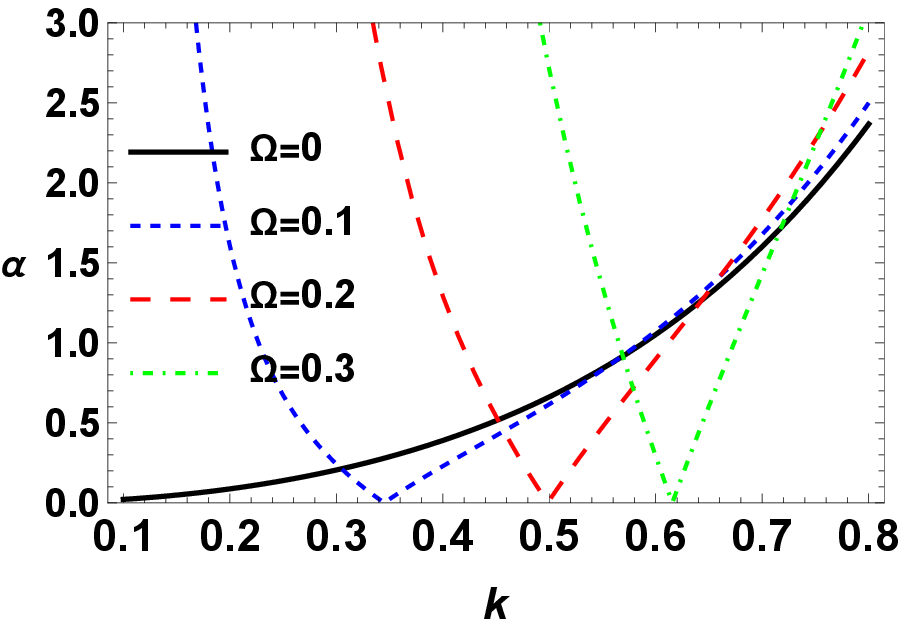}}
\caption{Plot of $\alpha$ vs. $k$ for different values of (a) $\delta_d$,  for fixed values of $\kappa_e=2$, $\Omega=0.2$; (b) $\kappa_e$, for fixed values of  $\Omega=0.2$, $\delta_d=0.2$; (c) $\Omega$, for fixed values of $\kappa_e=2$, $\delta_d=0.2$.}\label{f6}
\end{figure*}

\begin{figure*}
\centering
\subfigure[]{\includegraphics[width=2.5in]{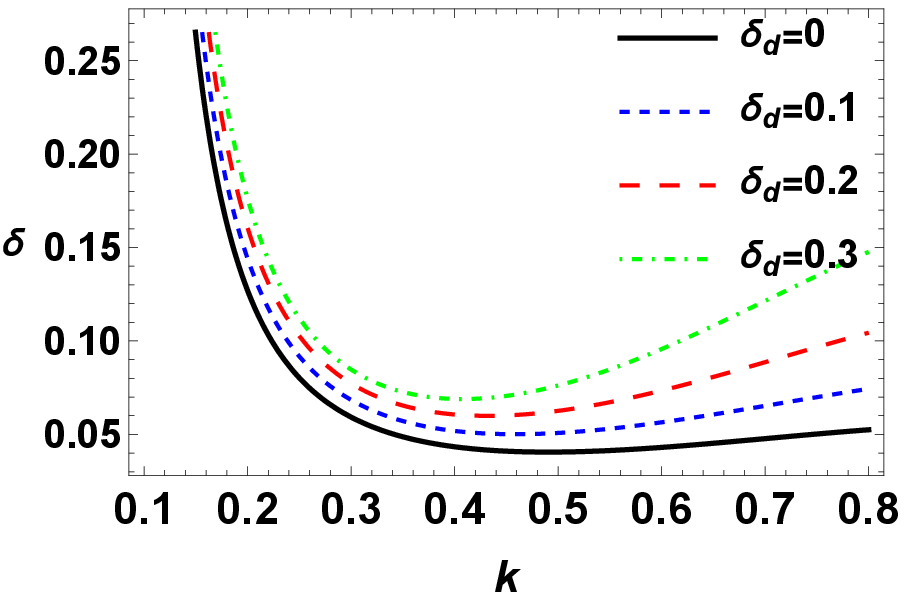}}
\subfigure[]{\includegraphics[width=2.5in]{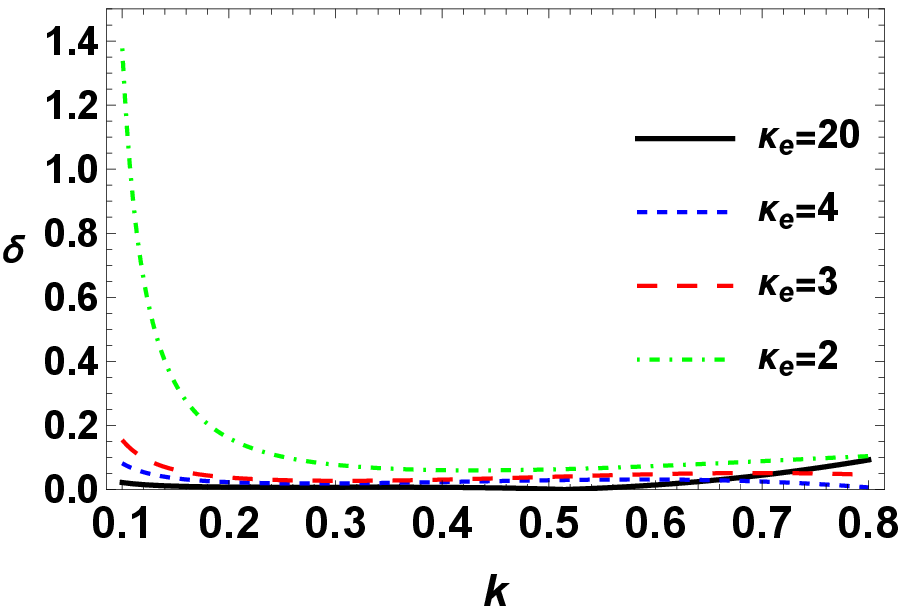}}
\subfigure[]{\includegraphics[width=2.5in]{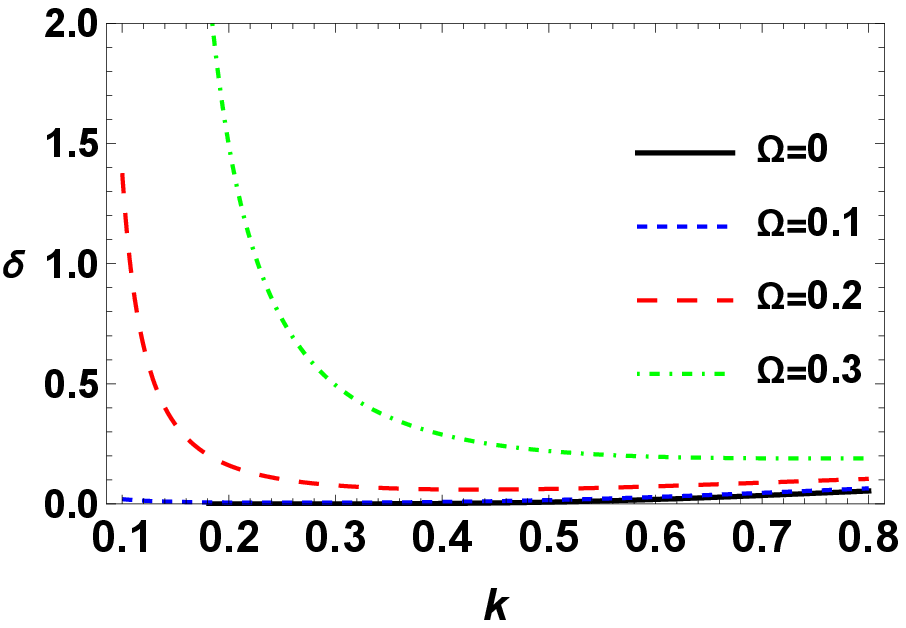}}
\caption{Plot of $\delta$ vs. $k$, for different values of (a) $\delta_d$,  for fixed values of $\kappa_e=2$, $\Omega=0.2$; (b) $\kappa_e$, for fixed values of  $\Omega=0.2$, $\delta_d=0.2$; (c) $\Omega$, for fixed values of $\kappa_e=2$, $\delta_d=0.2$.}\label{f7}
\end{figure*}

\begin{figure*}
\centering
\subfigure[]{\includegraphics[width=2.5in]{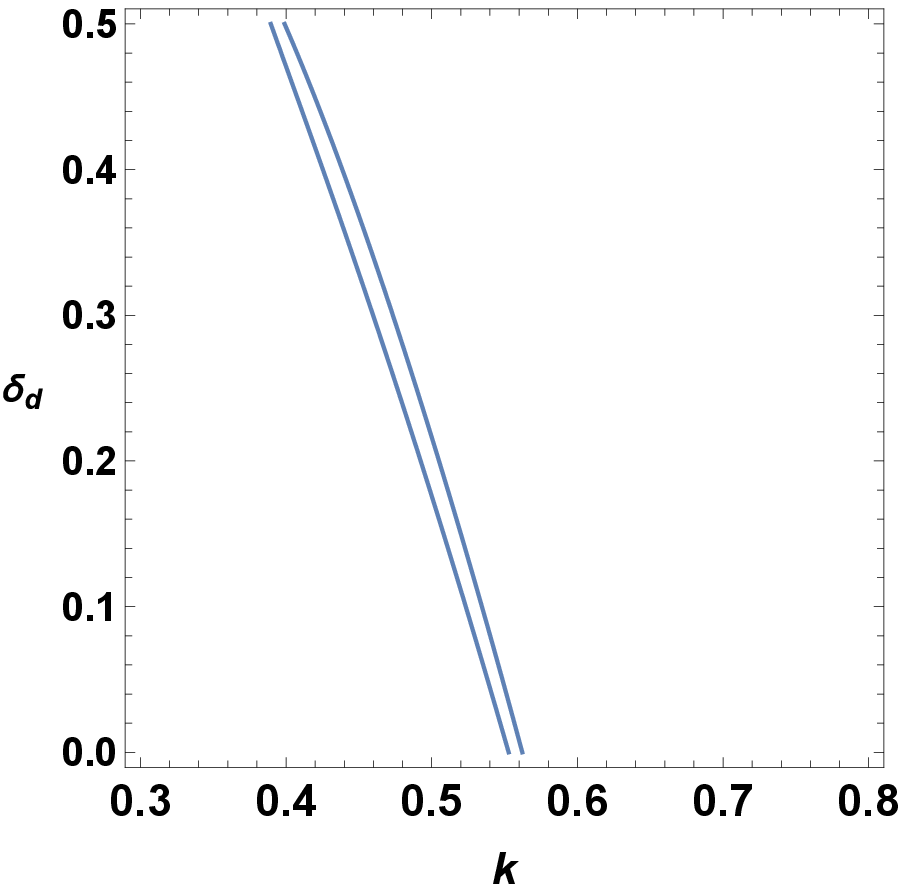}}
\subfigure[]{\includegraphics[width=2.5in]{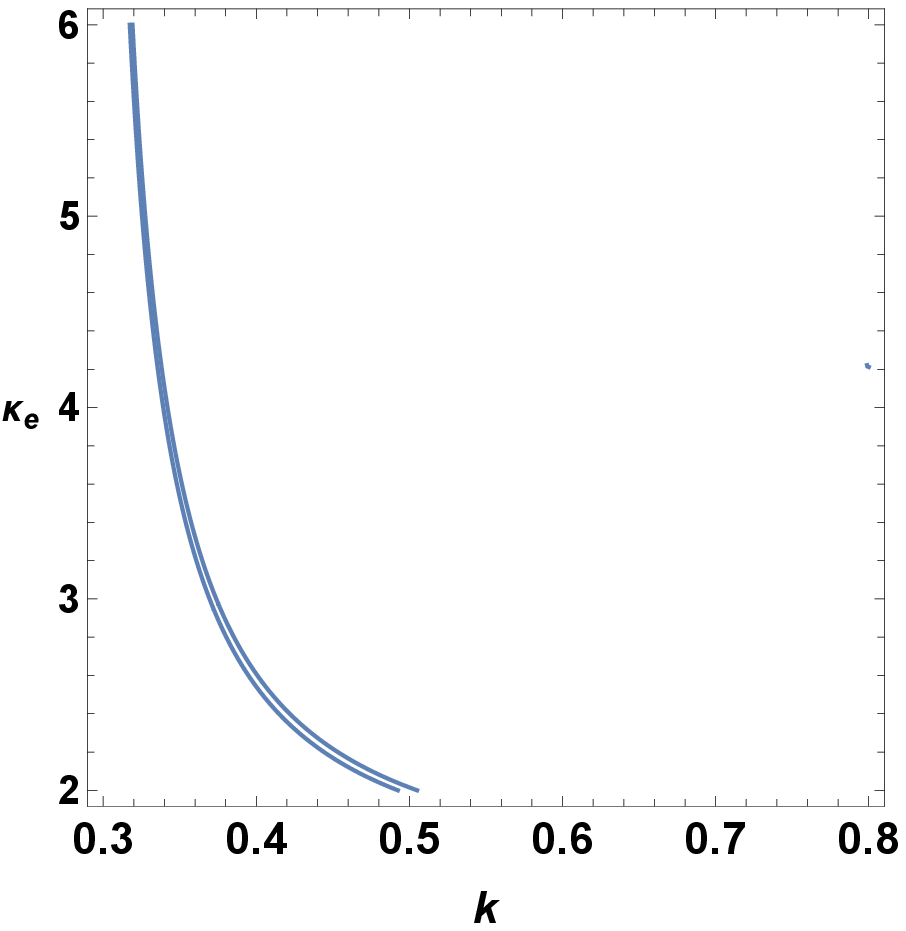}}
\subfigure[]{\includegraphics[width=2.5in]{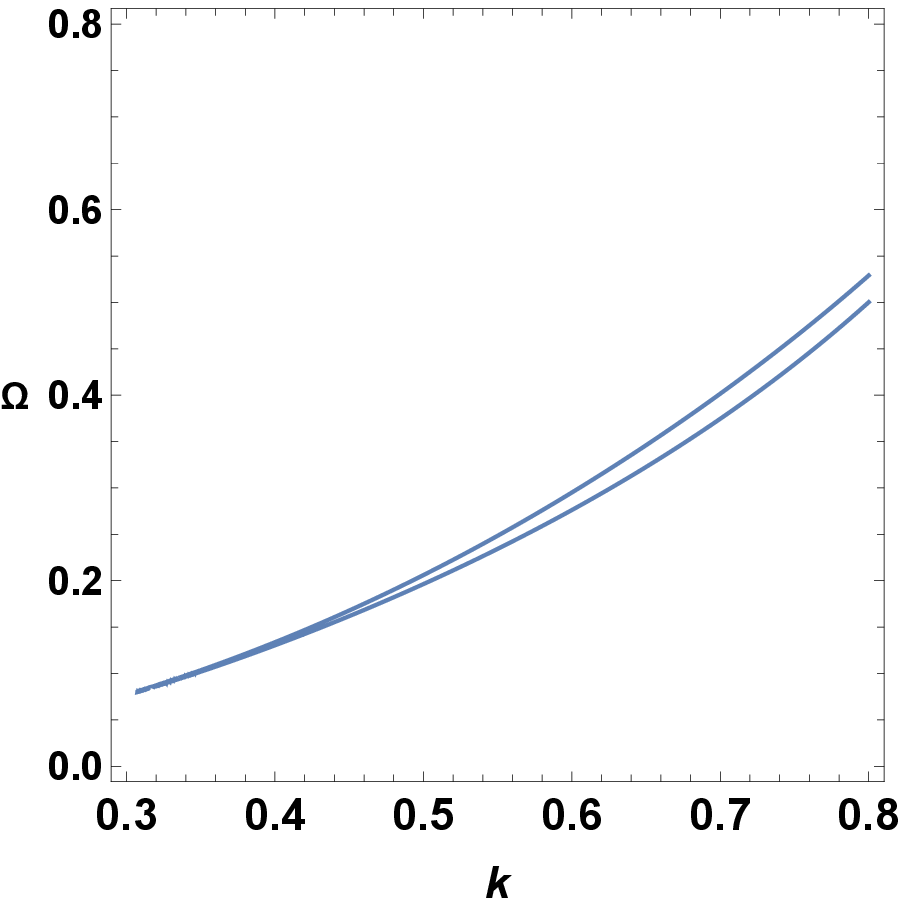}}
\caption{ContourPlot of $\frac{\alpha}{\delta}=1$ in the $k$-$\delta_d$ plane for fixed values of $\kappa_e=2$, $\Omega=0.2$; (b) $k$-$\kappa_e$ for fixed values of  $\Omega=0.2$, $\delta_d=0.2$; (c) $k$-$\Omega$ for fixed values of $\kappa_e=2$, $\delta_d=0.2$.}\label{f8}
\end{figure*}

\begin{figure*}
\centering
\subfigure[]{\includegraphics[width=2.5in]{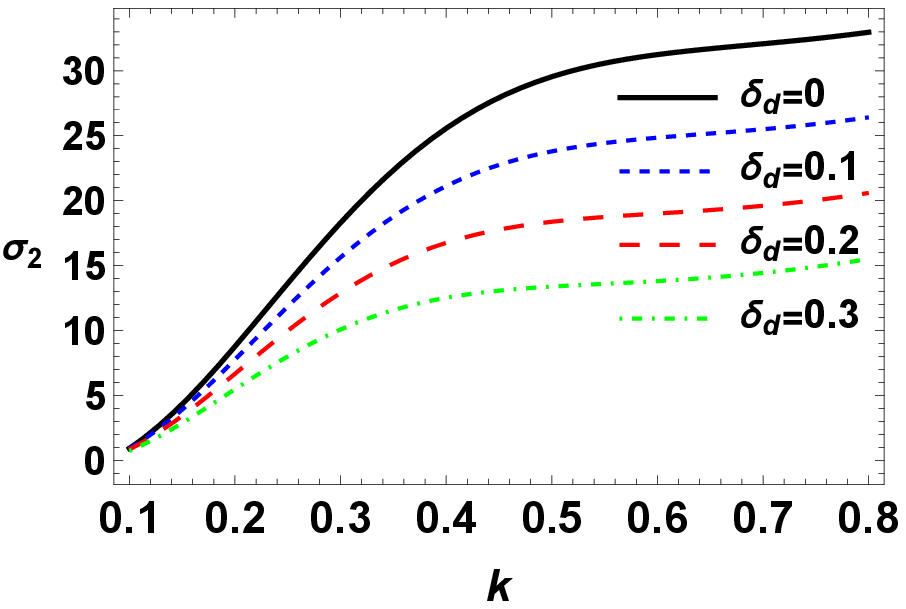}}
\subfigure[]{\includegraphics[width=2.5in]{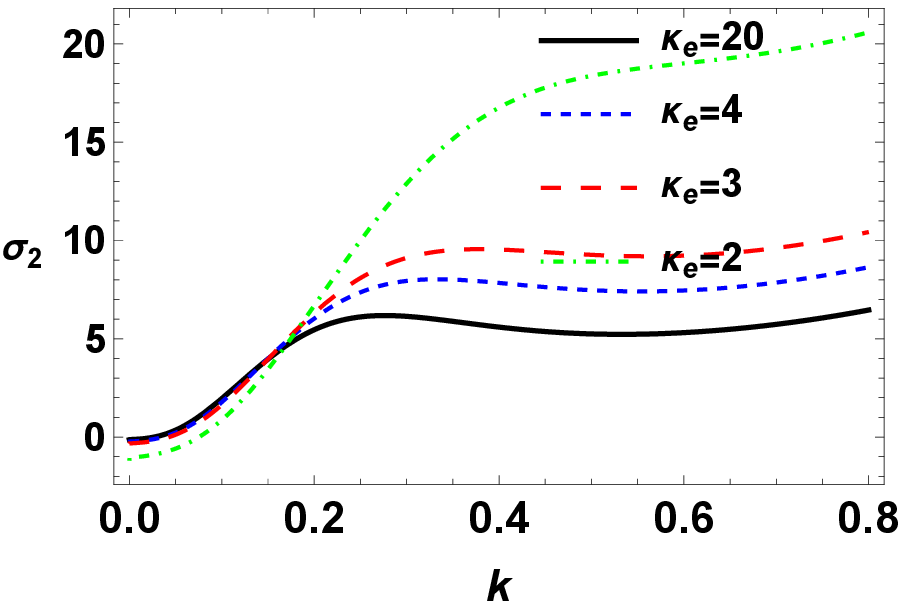}}
\subfigure[]{\includegraphics[width=2.5in]{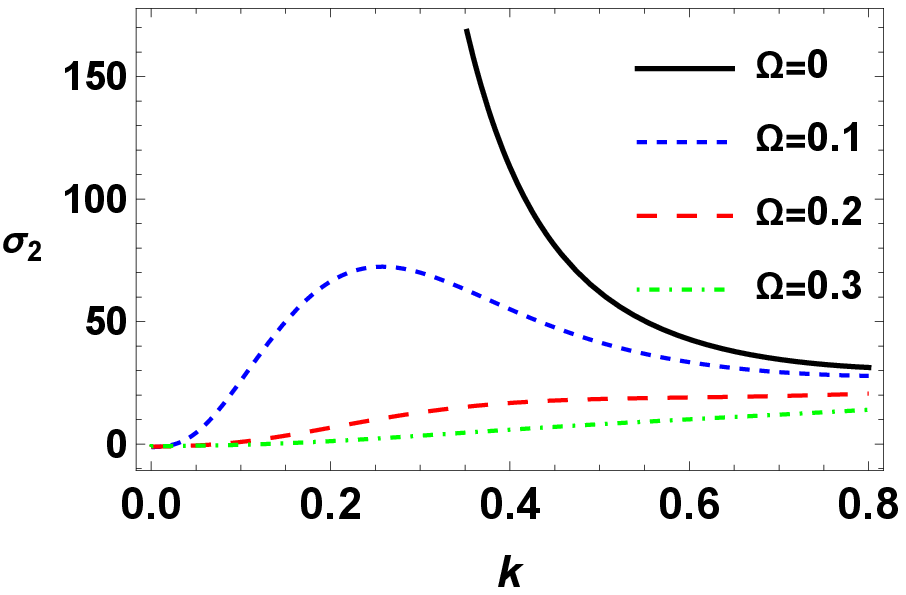}}
\caption{Contour plots of the quantity $\left(\frac{Q_2Q_3}{P_4}-Q_1\right)$ (whose sign in $\sigma_2$, defined in (29)) depicted versus $k$, for different values of (a) $\delta_d$,  for fixed values of $\kappa_e=2$, $\Omega=0.2$; (b) $\kappa_e$, for fixed values of  $\Omega=0.2$, $\delta_d=0.2$; (c) $\Omega$, for fixed values of $\kappa_e=2$, $\delta_d=0.2$.}\label{f9}
\end{figure*}

\begin{figure*}
\centering
\subfigure[]{\includegraphics[width=2.5in]{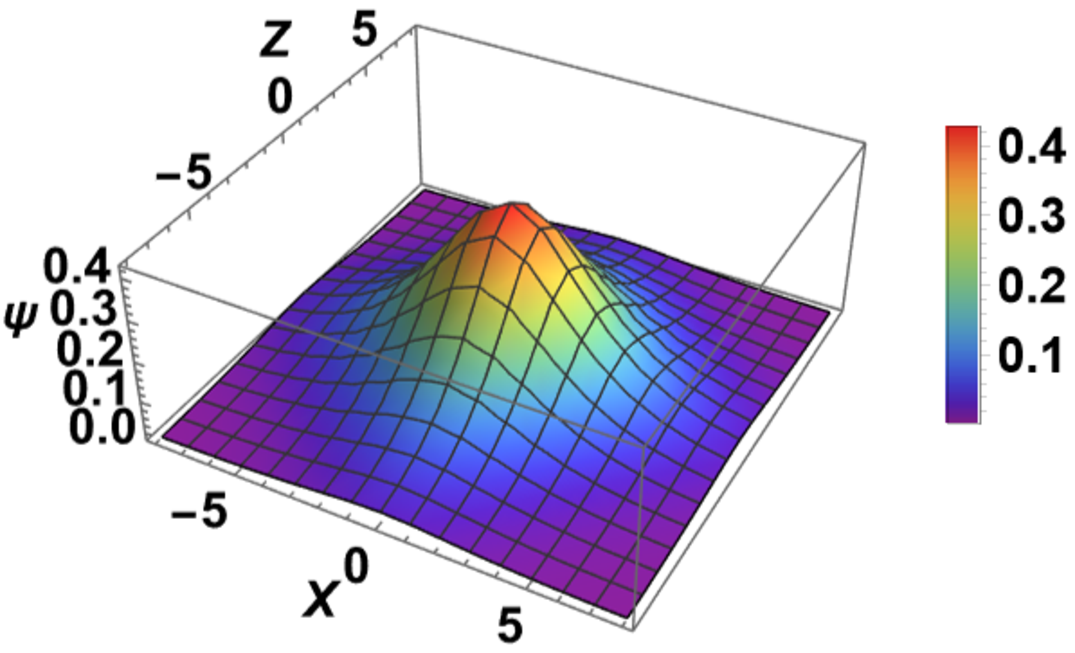}}
\subfigure[]{\includegraphics[width=2.5in]{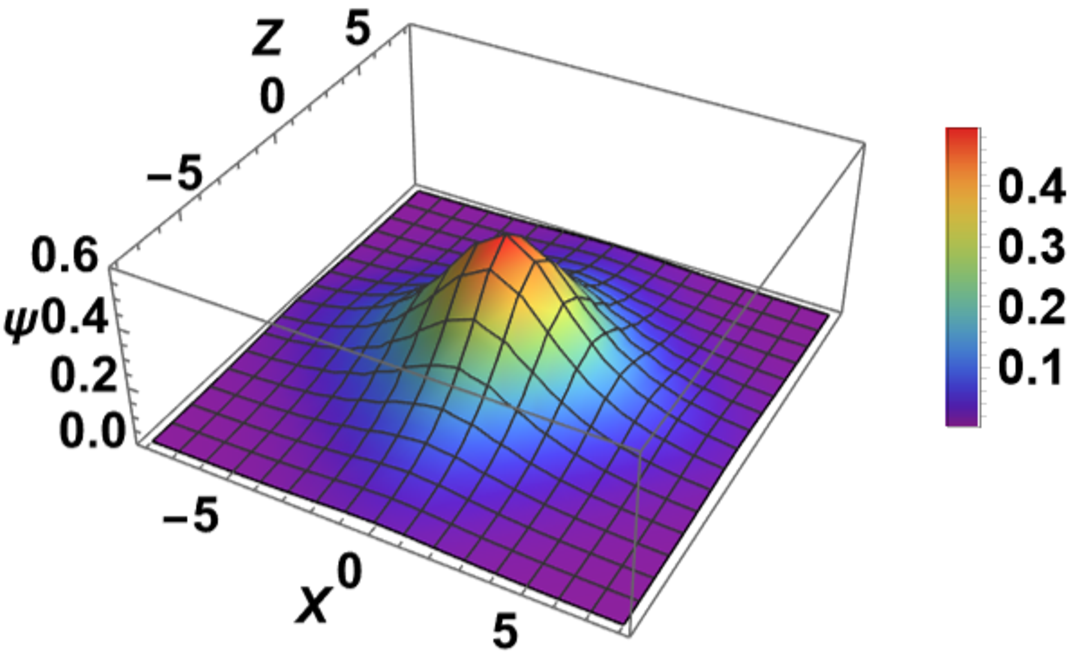}}
\caption{DIA dromion profile $\psi$ in the $X-Z$ plane, for (a) $\delta_d=0$; (b) $\delta_d=0.5$. Here, $\kappa_e=2$ and $\Omega=0.2$.}\label{f10}
\end{figure*}

\begin{figure*}
\centering
\subfigure[]{\includegraphics[width=2.5in]{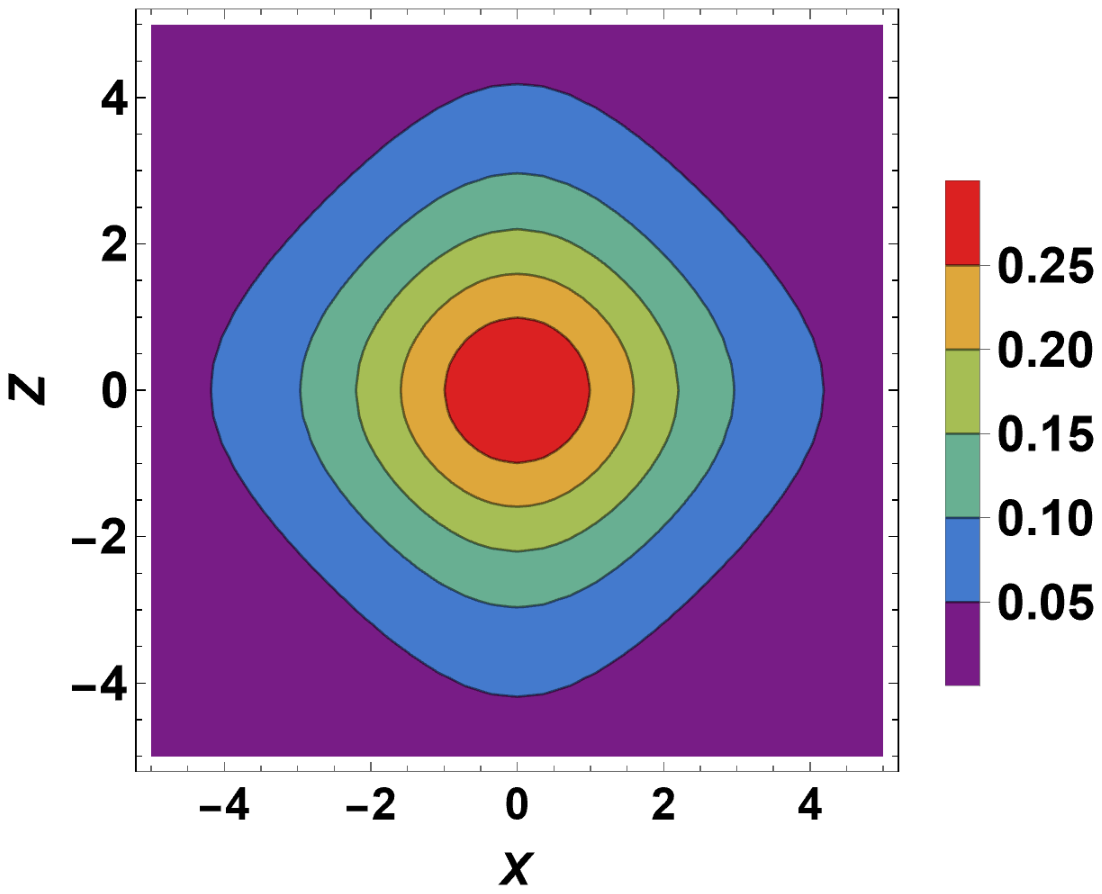}}
\subfigure[]{\includegraphics[width=2.5in]{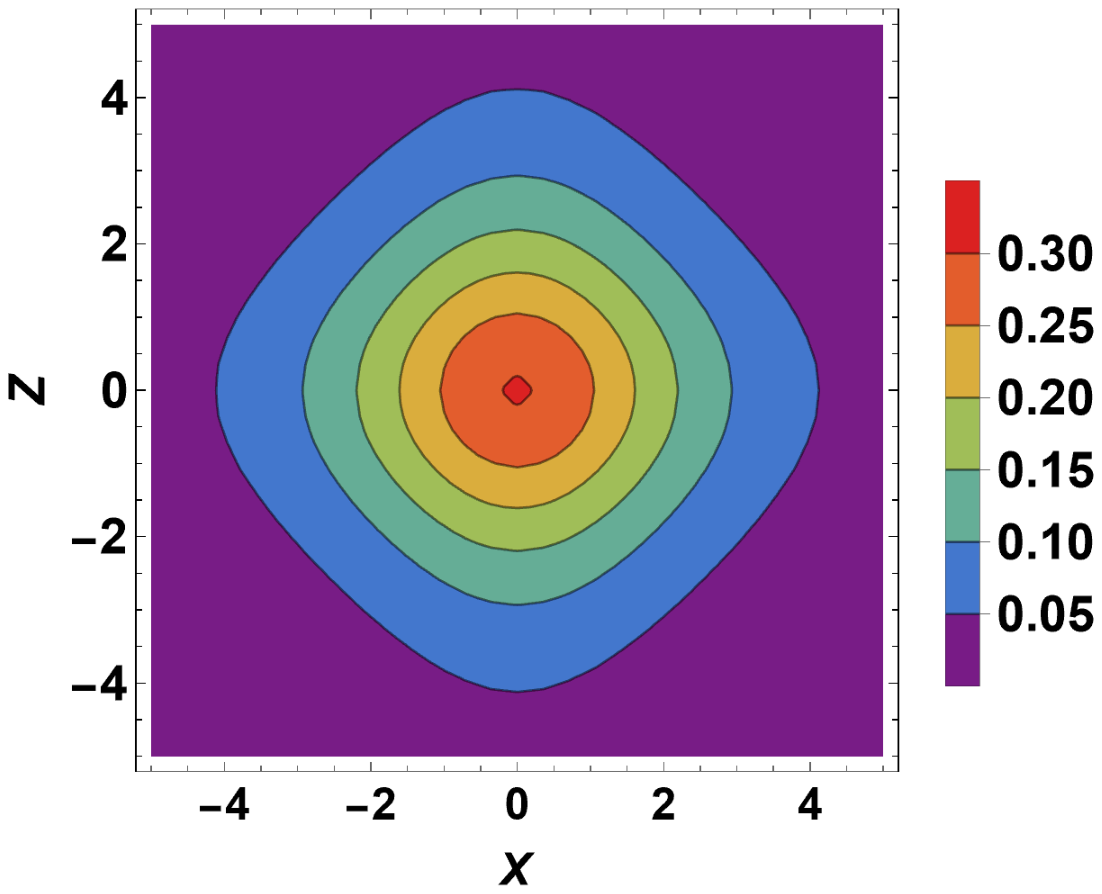}}
\caption{Contour map profile of a DIA dromion $\psi$ in the $X-Z$ Plane for (a) $\kappa_e=2$; (b) $\kappa_e=20$, for fixed values of $\delta_d=0.2$, $\Omega=0.2$.}\label{f11}
\end{figure*}

\begin{figure*}
\centering
\subfigure[]{\includegraphics[width=2.5in]{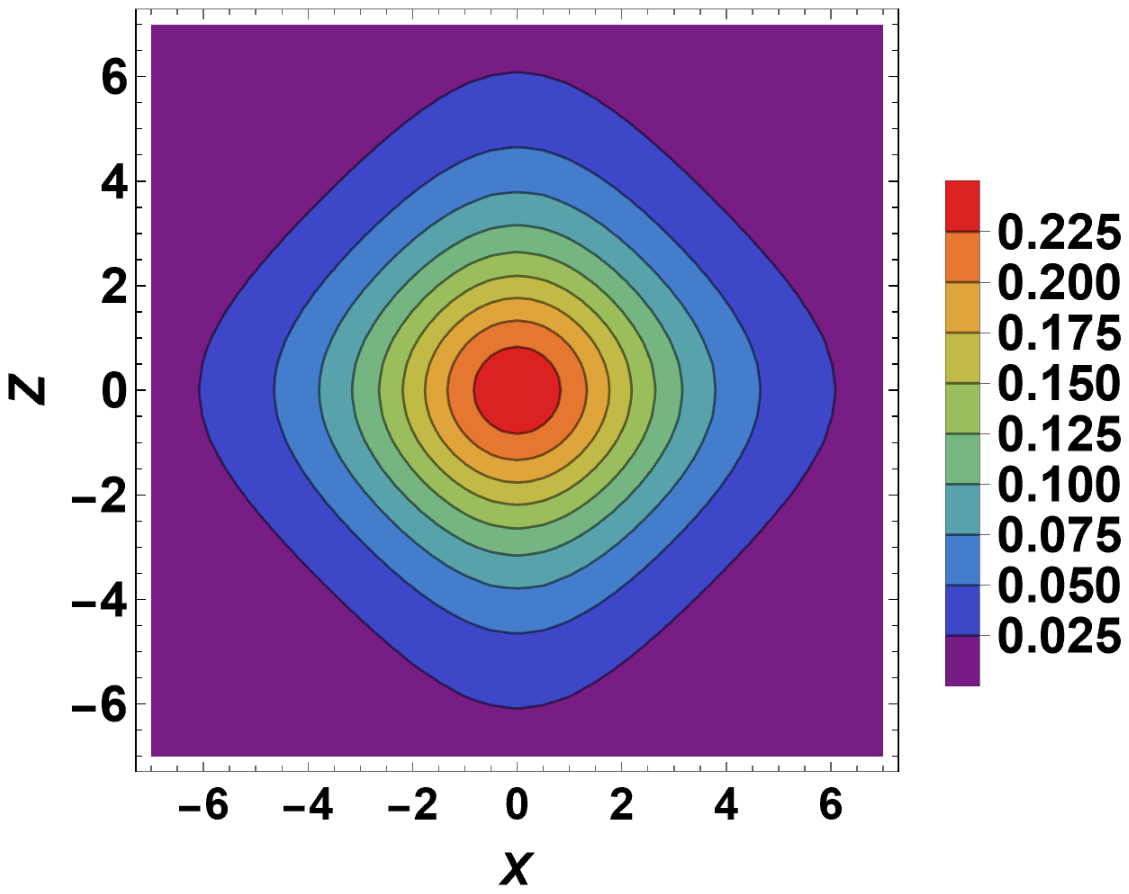}}
\subfigure[]{\includegraphics[width=2.5in]{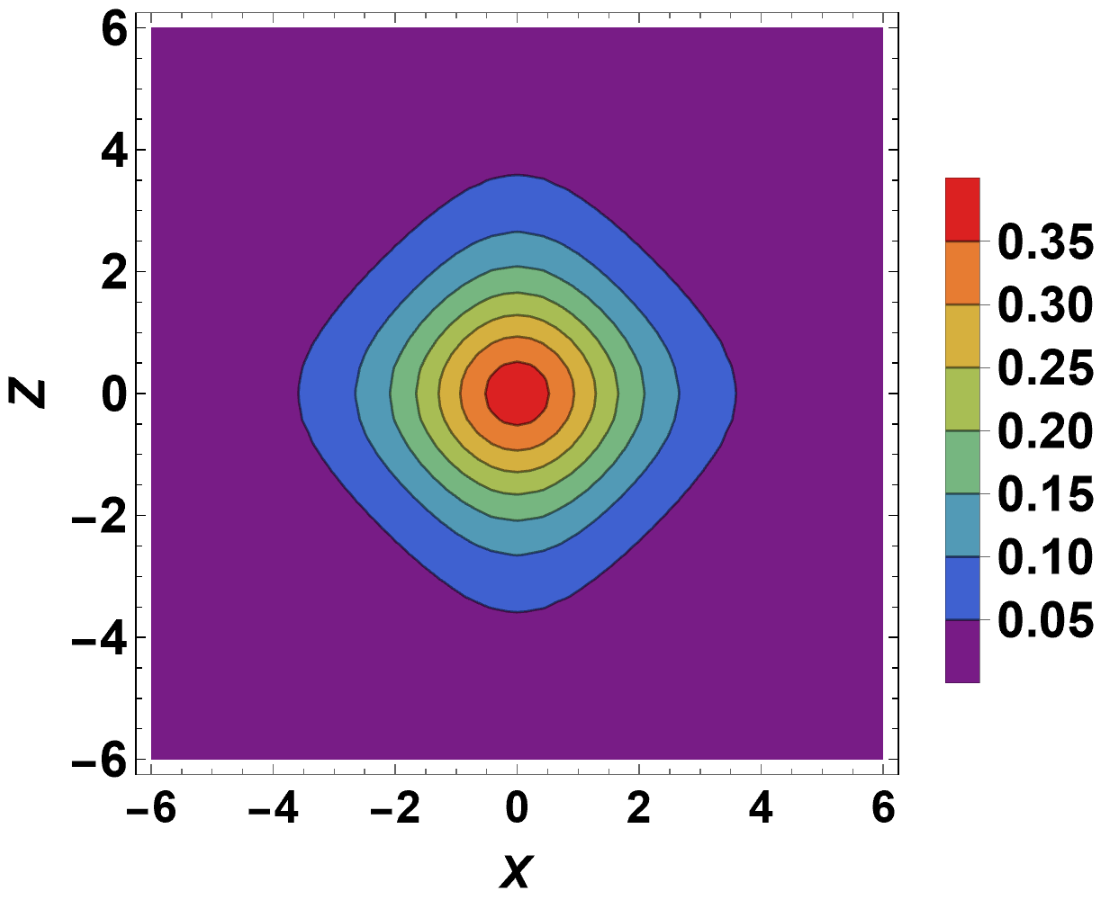}}
\caption{Contour map profile of a DIA dromion $\psi$ in the $X-Z$ Plane for (a) $\Omega=0$ (b) $\Omega=0.4$, for fixed values of $\delta_d=0.2$, $\kappa_e=2$.}\label{f12}
\end{figure*}

\subsection{Nonlinear analysis in 3rd order: amplitude modulation}

To simplify the algebra (and find a tractable relation among the zeroth harmonic terms), in the following we will consider $k_z=0$ (perpendicular propagation), so that $k=k_x$ will henceforth be assumed.  The carrier is thus assumed to propagate across the magnetic field. We are thus left with one branch in the dispersion relation, i.e. the upper branch $\omega_+$ (the only one that survives for $k_\parallel = 0$). As a consequence,  the group velocity has only one non-zero component, in fact the one   perpendicular to the magnetic field. \color{black} (Recall expressions (\ref{vg1})-(\ref{vg2}) above.) The wavenumber is thus expressed, in the chosen coordinate axes, as
\( \vec{k}=\left(k, 0, 0\right)\)
i.e., considering propagation along the x-axis. In what follows, therefore, the carrier is assumed to propagate in the perpendicular direction (across the magnetic field),   while the envelope propagates across the magnetic field;
hence, the cross-$\perp$-$\parallel$ spatial derivatives in all equations obtained to 2nd order will vanish.
(Note e.g. the discussion carried out in \citep{yashika}; cf. (33)-(36) therein, in full agreement with our consideration here.)  \color{black}  In what follows, therefore, the wavenumber (modulus) is essentially $k = k_\perp = k_x$, while $\omega$ is understood to be $\omega_+$  everywhere; also, in Figs. 3 and beyond.

In 3$^{rd}$ order in $\eps$, the condition for annihilation of secular terms leads to a closed system of equations in the form:
\ba&& i\frac{\pd\psi}{\pd T}+P_1\frac{\pd^2\psi}{\pd X^2}+P_2\frac{\pd^2\psi}{\pd Z^2}+Q_1\pss\psi+Q_2\psi \Phi=0\nn\\
&& P_3\frac{\pd^2 \Phi}{\pd X^2}+P_4\frac{\pd^2\Phi}{\pd Z^2}+Q_3\frac{\pd^2\pss}{\pd Z^2}=0  \label{dssystem} \ay
in terms of $\psi=\phi_{1}^{(1)}$ and $\Phi$.
 For clarity, note that the independent variables appearing in the latter system of equations are actually $\{X, Z, T\} = \{ X_1 - v_{g,x} T_1, Z_1, T_2\}$, i.e., the subscripts are dropped, for simplicity, in the algebra to follow.
All coefficients in the Davey Stewartson equation (DS) system above are real and defined in the Appendix.

The above system of evolution equations recovers the structure of the nonlinear Schr\"odinger (NLS) equation \citep{ik05}

\section{Classifying the Davey-Stewartson system}

In its general formulation, the Davey-Stewartson system of equations \citep{ds} can be divided into four main categories, based on the relative signs of $(P_1,P_2)$ and $(P_3,P_4)$.\\

(Case 1) \textit{The Elliptic-Elliptic Case:} If both  $P_1P_2$ and $P_3 P_4$ are positive,  then a solution in the form of a line soliton will exist, propagating along  $x$-axis with periodicity in $y$ \citep{Groves}. This is not the case in the particular model considered in this paper, since the group velocity is positive and has  negative curvature. \\

(Case 2) \textit{The Hyperbolic-Hyperbolic Case:} This occurs when both $P_1P_2$ and $P_3P_4$ are negative. According to \citet{Ghidaglia} this does not seem to arise naturally when working with water waves. Various solutions, including rogue waves, breathers, solitons and hybrids involving all three are known to exist in this case \citep{Rao}.\\

(Case 3) \textit{The Elliptic-Hyperbolic Case:} This happens for  $P_1 P_2>0$ and $P_3P_4<0$ and often known as DS-I system. Davey and Stewartson were the first to illustrate this algebraic system of equations in their investigation of water waves \citep{ds}. These equations can be explicitly solved by using either inverse scattering method \citep{Fokas} or  Hirota's Bilinear method \citep{Satsuma}. In summary, earlier investigations have led to the following conslusions: (i) by imposing  time-independent boundary conditions, the initial perturbation will bifurcate into a number of 2D breathers. Likewise, (ii) adopting time-dependent boundary conditions, an initial disturbance will bifurcate into a number of 2D localized  travelling structures. 2D localized solutions  correspond to the discrete spectrum of bound states, which imply that they are nonlinear distortions. Unlike 1D solitons, these 2D  solutions do not  conserve their shape and size during  interaction or exchange energy \citep{Fokas}. In a different approach, \citet{White} adopted necessary non-trivial boundary conditions on $\Phi$ in order to unveil soliton or dromion solutions analytically. 
\\

(Case 4) \textit{The Hyperbolic-Elliptic Case:} This case is realized only when $P_1 P_2 < 0$ and $P_3 P_4 > 0$ which means that no dromion structures will form in this regime. \citet{McConnell} and \citep{Klein} have carried out an analytical investigation of singular, hump and freak wave type solutions. Freak waves were examined explicitly by employing Hirota's bilinear method \citep{Ohta}. \citet{Kavitha} determined a solution in exponential form for the standard DS system. \\

Earlier works have shown that the latter  (hyperbolic-elliptic) case, also known as the DS-II system, occurs in unmagnetized plasmas, in relation with ion-acoustic \citep{Nishinari} and dust-ion acoustic \citep{xue04,Singh2022} waves. On the other hand, the DS-I (elliptic-hyperbolic) regime occurs in magnetized plasmas, in association with dust-acoustic \citep{aDuan,yashika} or electron-acoustic (Langmuir) \citep{Ghosh} waves. \\

Our model is covered by \textit{Case 3} above (i.e. is of DS-I type)  predominantly, for small values of the carrier wavenumber $k$ (i.e. for long carrier wavelength), in fact depending parametrically on the plasma properties (via the values of $\delta_e$ and $\kappa_e$). However, the system's behavior may turn to \textit{Case 2} for larger wavenumbers, say $k \gtrsim 0.35$ for $\Omega=0.1$: see in Figure  \ref{f4} (depending on plasma parameters), where the various curves mark this transition from one regime to the other. (In Figure 6, the regions on the left of the curves belong to \textit{Case 3}, while the regions on the right are in \textit{Case 2}.) Note that the product $P_3 P_4$ comes out to be negative always, in our case; cf. Fig. \ref{f5}. On the other hand, for unmagnetized plasma, only \textit{Case 4} (DS-II) is valid \citep{xue04,Singh2022}. (As a matter of fact, the coefficients are different in the unmagnetized plasma case; for instance, $P_3 > 0$ then \citep{Singh2022}, whereas it is negative in the magnetized case, as seen above. Also, note that the  product $P_3 P_4 = -1/c_1 < 0$ in the present case, unlike in unmagnetized plasma.)  Note, for completeness, that \textit{Case 1} (elliptic-elliptic) is never realized in the fluid-plasma framework adopted in our work.

\section{Dromion solutions: analytical setting}

Dromions are localized, lump-like solutions that, unlike solitons, can scatter inelastically, which is to say they can exchange energy. This is in contrast to 1D solitons, which undergo a complicated interaction over some finite interval in time, but then their asymptotic states post-collision differ only by a phase-shift. They arise in a very particular case of the DS-I equations, with certain boundary conditions \citep{ghosh02}. To derive a dromion-type solution of the DS system above, we begin by  re-scaling the system of equations (\ref{dssystem})  by using  transformation as $\Phi\mapsto\ \Phi-Q_3/P_4|\psi|^2$ \citep{Hietarinta90}.
\ba&& i\frac{\pd\psi}{\pd T}+P_1\frac{\pd^2\psi}{\pd X^2}+P_2\frac{\pd^2\psi}{\pd Z^2}+\left(Q_1-\frac{Q_2Q_3}{P_4}\right)\pss\psi+Q_2\psi \Phi=0\nn\\
&& P_3\frac{\pd^2\Phi}{\pd X^2}+P_4\frac{\pd^2\Phi}{\pd Z^2}-\frac{Q_3P_3}{P_4}\frac{\pd^2\pss}{\pd X^2}=0\label{ds} \, . \ay

Then, the above equations are further rescaled (except at $k=0$) as described in the following.
By scaling  $x\mapsto x/\sqrt{|P_1|}$,  $y\mapsto y/\sqrt{|P_2|}$, $\psi\mapsto \psi\sqrt{|Q_1-Q_2Q_3/P_4|}$ and $\Phi\mapsto \Phi|Q_2|$ and defining
\ba\sigma_{1}=-\text{sgn}(P_1);\quad\sigma_2= \text{sgn}\left(\frac{Q_2Q_3}{P_4}-Q_1\right)\nn \\
\sigma_3= \text{sgn}(Q_2); \qquad \sigma_4= \text{sgn}(P_3) \, , \ay

we obtain
\ba&& i\frac{\pd\psi}{\pd t}-\sigma_1\frac{\pd^2\psi}{\pd x^2}+\frac{\pd^2\psi}{\pd y^2}-\sigma_2\pss\psi+2\sigma_3\psi \Phi=0\\
&& \sigma_4\frac{\pd^2\Phi}{\pd x^2}+\alpha\frac{\pd^2\Phi}{\pd y^2}+\delta\frac{\pd^2\pss}{\pd x^2}=0\label{dseqn} \, . \ay
Note the appearance of a factor 2 beside $\sigma_3$ as a deliberate choice in the scaling, to aid comparison between the calculations which follow and the work published in \citep{Satsuma} as a basis.

The effects of the various coefficients derived in the previous section can be bundled into two terms:
\ba \alpha &=&\frac{|P_1|P_4}{|P_2P_3|}\nn\\
\delta&=&-\frac{sgn(P_3)|Q_2|Q_3}{|Q_2Q_3-P_4Q_1|}\label{alphadelta}\ay

Dromion solutions can be constructed by Hirota's method. This is a particularly restrictive form of solution as it places constraints on the values of $\alpha$ and $\delta$ as well as on the forms of $\psi$ and $\Phi$. We will use the \textit{Ansatz}
\be \Phi=2\sigma_2\delta \left(\log f\right)_{ZZ};\quad\psi=\frac{g}{f}; \qquad  f\in\mathbb{R}\e
such that $\sigma_4=\sigma_1=-1$, $\alpha=1$ and $\delta=\pm 1$ to obtain dromions. Therefore,    Eqs. (\ref{dseqn}) become
\ba \left(iD_T+D_X^2+D_Z^2\right)g\cdot f&=&0\nn\\
\left(- D_X^2+  D_Z^2 \right)f\cdot f&=&-\sigma_2 g.g^{\ast} \, \ay
where $D_{X,Z}$ are the Hirota bilinear operators. This is simplified by rotating the coordinate system by $\pi/4$.
\ba \left(iD_T+D_X^2+D_Z^2\right)g\cdot f&=&0\nn\\
2D_{XZ}f\cdot f&=&\sigma_2g.g^{\ast}\label{rotds}\ay
Now, expanding the new variables as
\ba g= \epsilon g^{(1)}+ \epsilon^3 g^{(3)}+ ...,\nn\\
f=1+\epsilon^2 f^{(2)}+ \epsilon^4 f^{(4)}+ ...,\ay
where $\epsilon$ is a small quantity and equating for different orders we obtain
\begin{equation} ig_{T}^{(1)}+ g_{XX}^{(1)}+g_{ZZ}^{(1)}\label{con1} \, , \end{equation}
\begin{equation} f_{XZ}^{(2)} = g^{(1)}g^{\ast(1)}\label{con2} \, \end{equation}
and so on. Solving Eq. (\ref{con1}) we get
\be g^{(1)}=\sum_{j=1}^{N} \exp(\chi_j), \quad  \chi_j= p_j X+s_j Z-i\omega T +b_j \label{con3}\e
where $p_j$, $s_j$ and $b_j$ are complex constants satisfying the condition
\be p_{j}^2+s_{j}^2-\omega=0\label{con31} \, . \e

In order to construct a one-dromion solution, we may fix $N=1$. Put Eq. (\ref{con3}) in Eq. (\ref{con1}) we get
\be f^{(2)} =\exp(\chi_1+\chi_{1}^{\ast}+2\vartheta), \qquad \exp(2\vartheta)=\frac{\sigma_2}{16p_{1R}s_{1R}}\label{con4}\e
where $p_{1R}$ and $s_{1R}$ are real parts only. Equating higher orders of $\epsilon$, one can get the $g^{(j)}=0$ for $j\geq3$ and $f^{(j)}=0$ for $j\geq4$, respectively \citep{ghosh02}. To determine an analytic form for a dromion solution, we employ ansatz
\be f=1+\exp(\chi_1+\chi_1^{\ast})+\exp(\chi_2+\chi_2^{\ast})+\exp(\chi_1+\chi_1^{\ast}+\chi_2+\chi_2^{\ast}),\label{con5}\e
where $\chi_1=p_1 X+ip_{1}^2T+b_1$ and $\chi_2=s_1 Z+is_{1}^2T+b_2$. Substituting Eq. (\ref{con5}) in  Eq. (\ref{con2}), we get
\be g=4\sqrt{\sigma_2 p_{1R}s_{1R}}\exp(\chi_1+\chi_2)\label{con6} \, .\e
For simplicity, we have assumed that $p_1=s_1=p$ and $T=0$. The dromion solution thus obtained is given by
\be \psi(X,Z,T=0)=\frac{4p\sqrt{\sigma_2}\exp{p(X+Z)}}{1+\exp{2p X}+\exp{2p Z}+\exp{2p(X+Z)}} \, \e
Here, $p=\sqrt{\omega/2}$ from Eq.(\ref{con31}). Noted that $\sigma_2>0$.

\begin{table*}
\caption{An indicative set of wavenumber ($k$) component values for $\frac{\alpha}{\delta}=1$ are listed, i.e. as required to fulfill the necessary condition for the existence of dromions.}
\centering

\begin{tabular}
{ p{0.04\linewidth}|p{0.12\linewidth}|p{0.12\linewidth}|p{0.12\linewidth}|p{0.12\linewidth}|p{0.12\linewidth}}
\hline
\hline

S.no & $\delta_d$ & $\kappa_e$ & $\Omega$ & $k_1$ & $k_2$ \\
\hline

\multirow{6}{*}{1}  &  0  &    &    & 0.5527 & 0.5631  \\
                    & 0.1 &    &    &  0.5233 & 0.5344 \\
                    & 0.2 & 2  & 0.2 & 0.4924 & 0.5049  \\
                    & 0.3 &    &     & 0.4597 & 0.4733  \\
                    & 0.4 &    &    &  0.4252 & 0.4386 \\
                    & 0.5 &    &    & 0.3895 & 0.3989 \\   \hline
\multirow{6}{*}{2}  &     &  2 &     & 0.4924 & 0.5049  \\
                    &     &  3 &    &  0.3694 & 0.3738 \\
                    & 0.2 &  4 & 0.2 & 0.3392 & 0.3420  \\
                    &     &  5 &    & 0.3274  & 0.3253  \\
                    &     &  6 &    &  0.3173 & 0.3191 \\
                    &     &  20&    & 0.2951  & 0.2959 \\  \hline
\multirow{6}{*}{3}  &     &    & 0    & --& --  \\
                    &     &   &  0.1  &  0.3436 & 0.3464 \\
                    & 0.2 &  2 & 0.2  & 0.4929 & 0.5049  \\
                    &     &   &  0.3  & 0.6050  & 0.6263  \\
                    &     &   &  0.4  &  0.6986 & 0.7224 \\
                    &     &  &   0.5  & 0.7788  & 0.7998 \\   \hline
\hline
\end{tabular}\label{t1}
\end{table*}

\section{Parametric analysis}

In this section, we shall present the results of numerical analysis that we have performed, to examine
the existence conditions of DIA dromions in magnetized non-Maxwellian dusty plasma. To this end, we have adopted a data set corresponding to read data from Saturn's magnetosphere: $n_{i, 0}=  1-20 cm^{-3}$, $n_{e0} = 2-45 cm^{-3}$, $Z_d = 10^3$, $n_d = 10^{-3}-10^{-1} cm^{-3}$, $T_e = 1-10$ eV \citep{yar07}. Our parametric investigation was  based on these values, used a basis to compute the numerical values of the coefficients of the DS system, modelin the evolution of DIA dromions.

In order to gain insight into the impact of various parameters on the dispersion characteristics of DIA waves, the dispersion curves have been depicted in Fig. \ref{fex1} and \ref{fex2} for different values of the spectral index, in fact for both magnetized  and unmagnetized plasma (for comparison).

\subsection{Parametric dependence on the dust concentration}

In Fig. \ref{f1}(a), we have depicted the variation of the angular frequency $\omega$ vs. the wavenumber $k$, for different values of $\delta_d$ (dust concentration). It is obvious that both the frequency and the phase speed of DIA wavepackets increase with stronger dust presence. This behavior is reflected in the group velocity (see \ref{f1}(b)), which actually increases with the dust (and remains positive everywhere).
The $P_1$ coefficient, however (depicted in Fig. \ref{f1}(c)) presents a more perplex behavior, as it increases for higher $\delta_d$ for small $k$, but then starts to decrease for higher $\delta_d$ after some inflection point, and eventually changes sign (becomes negative) around $k=0.5$. (Note that this fact, among others, enables a switch between the DS-I and DS-II regimes, as discussed above).

The Z-dispersion coefficient $P_2$ -- shown in Fig. \ref{f1}(d) -- remains positive, and in fact increases with an increase in dust concentration (for larger $\delta_d$).

Fig. \ref{f1}(e) shows that $P_3$ remains negative and in fact increases (in absolute value) with an increase in $\delta_d$. The same trend is followed (in absolute value) by  $P_4$ ($> 0$): see Fig. \ref{f1}(e).

The analogous variation of the nonlinear coefficients $Q_{1, 2, 3}$ for different values of $\delta_d$ is shown in  Fig. \ref{f1}(g-i), respectively: all of these coefficients present a strong dependence on the value of  $\delta_d$ (i.e., dust concentration). Notice that $Q_3 > Q_2 > 0 $ everywhere, while $Q_1$ changes sign at a small value (say $k_0$) of the wavenumber $k$.

\subsection{Parametric dependence on electron superthermality}

Fig. \ref{f2}(a) depicts the variation of $\omega$ vs. $k$, for different values of $\kappa_e$. It is obvious that both the frequency and the phase speed increase with higher  $\kappa_e$ (or will decrease with a stronger deviation from thermal distribution, i.e. with a stronger suprathermal particle component). Fig. \ref{f2}(b) shows that the group velocity increases for lower values of $k$ and then it decreases after $k=0.5$. The group velocity  also decreases for lower $\kappa_e$ (i.e., strongly superthermal case). Fig. \ref{f2}(c) represents that  $P_1$  decreases for lower values of $\kappa_e$  and become negative around $k=0.4$. Similarly, Fig. \ref{f2}(d) shows that $P_2$  decreases  for lower $\kappa_e$. Fig. \ref{f2}(e) illustrates the variation of the $P_3$ coefficient for different values of $\kappa_e$. We notice that $P_3$ is negative and in fact increases (i.e. decreases in absolute value) for smaller $\kappa_e$. Fig. \ref{f2}(f) shows that $P_4$  is positive and actually decreases for lower $\kappa_e$.

Moreover, Fig. \ref{f2}(g) shows the variation of the $Q_1$ coefficient for different values of $\kappa_e$: we notice that $Q_1$  is positive below a certain (small) value of $k$ (say, a root $k_0\approx 0.1$), and becomes negative for $k > k_0$, and it decreases (i.e. increases in absolute value, in the negative region) with lower $\kappa_e$ (i.e., with more superthermal electrons).

In a similar manner, Fig. \ref{f2}(h-i) shows the variation of the $Q_{2}$ and $Q_{3}$ coefficients for different values of $\kappa_e$. One notices that $Q_2$ increases with  smaller $\kappa_e$ and attains a root at some value of the wavenumber k, beyond which it becomes negative. (The root of $Q_2$ is a decreasing function of $\kappa_e$.) $Q_3$, on the other hand, appears to be positive everywhere.

Fig. \ref{f3}(a) depicts $\omega$ versus $k$, for different values of $\Omega$. It is obvious that both the frequency and the phase speed of DIA wavepackets increase with a stronger magnetic field (i.e., for higher ion cyclotron frequency $\Omega$). Fig. \ref{f3}(b) shows that the group velocity  decreases with an increase in the magnetic field strength (via $\Omega$). Fig. \ref{f3}(c) illustrates that $P_1$  is negative for $\Omega=0$ (unmagnetized case) \citep{Singh2022}  but becomes positive in the magnetized case, in the low $k$ range. Note that it decreases with higher $\Omega$.  $P_2$  also decreases with higher $\Omega$, as seen in Fig. \ref{f3}(d). It is important to mention here that $P_3$ is negative and in fact decreases (i.e. increases in absolute value) with higher $\Omega$, as obvious in Fig. \ref{f3}(e). In contrast with this, $P_4$  is always positive and increases with higher $\Omega$. As a consequence, the product $P_3 P_4$ is always negative in our case (for finite magnetic field).

\subsection{Parametric dependence on the magnetic field (strength)}

In Fig. \ref{f3}(g), one sees that $Q_1$  decreases with higher $\Omega$, while $Q_2$ is essentially independent of $\Omega$ --- see Fig. \ref{f3}(h).
Fig. \ref{f3}(i) shows that $Q_3$  increases with higher $\Omega$.

Fig. \ref{f4}(a) shows contours of the product $P_1 P_2$ on the $k - \kappa_e$ plane, for different values of $\delta_d$. The product $P_1 P_2$ is positive only for low values of the carrier wavenumber $k$ (long wavelength), while it becomes negative for higher $k$ (shorter wavelength). In the presence of dust, the region of positive  values of this product becomes narrower.

Similarly,  Fig. \ref{f4}(b) shows contours  of the product $P_1 P_2$ in the $k - \delta_d$ plane, for different values of $\kappa_e$. The product $P_1 P_2$  is positive only for low $k$ values, above which it becomes negative. In the presence of highly superthermal electrons, the region where $P_1 P_2$  is positive  becomes wider.

Moreover, Fig. \ref{f4}(c) shows contours of $P_1 P_2$ in the $k-\Omega$ plane, for different values of $\delta_d$. It is seen that, in the presence of magnetic field, the region of positive  values of product becomes narrower.

Fig. \ref{f5} (a-c) show that the product $P_3 P_4$ is always negative for the given values of plasma parameters.
Hence, our plasma model is described as a DS-I type (i.e., elliptic-hyperbolic) system of equations,  for small values of $k$. Therefore, dromions may occur in the given plasma model, for small values of the carrier wavenumber $k$.

\section{Dromion solutions: numerical results}

We recall that dromion existence relies on the quantities $\alpha$ and $\delta$, defined in Eq. (\ref{alphadelta}) above.
Fig. \ref{f6}(a-c) shows the variation of $\alpha$,  for different values of $\delta_d$, $\kappa_e$ and $\Omega$. Similarly,  Fig. \ref{f7}(a-c) shows the variation of $\delta$, for different values of $\delta_d$, $\kappa_e$ and $\Omega$.

Fig. \ref{f8}(a-c) shows contour plots (curves) along which ${\alpha}/{\delta} = 1$, for different values of $\delta_d$, $\kappa_e$ and $\Omega$; based on these controus, one can obtain the specific values of $k$, for the existence of dromions. Based on this information, we have chosen to isolate two k-values to fulfill the necessary conditions i.e., $\alpha=1$, $\delta=\pm 1$ one one hand, or $\frac{\alpha}{\delta}=1$ on the other. The numerical values of $k$ where the existence of dromions is possible, for different values of $\delta_d$, $\kappa_e$ and $\Omega$, are listed in Table \ref{t1}.

Fig. \ref{f9}(a-c) shows the variation of the quantity $\left(\frac{Q_2Q_3}{P_4}-Q_1\right)$, whose sign is  $\sigma_2$, defined in (29) above, for different values of $\delta_d$, $\kappa_e$ and $\Omega$. It turns out that this is a positive quantity, viz. $\sigma_2 = +1$ in the given  parameter range of values.

We have computed the dromion solutions corresponding to certain parameter values satisfying the above constraints.
The obtained lump-shaped dromion profile is depicted in Fig. \ref{f10} (a,b), for different values of $\delta_d$. A numerical investigation reveals that the dromion amplitude increases with higher dust concentration (i.e. with larger $\delta_d$).

Fig. \ref{f11} (a,b) shows contour plots of DIA dromions for different values of $\kappa_e$. The dromion amplitude is seen to decrease for lower $\kappa_e$ (i.e., with an increase in the superthermal electron population component).

Fig. \ref{f12} (a,b) shows contour plots of DIA dromion solutions, for different values of magnetic field strength (via $\Omega$). It is observed that the amplitude (width) of these solutions increases (decreases) with higher $\Omega$. The stronger the magnetic field (strength), therefore, the higher (amplitude) and narrower (in width) dromions will be.

Concluding, the plasma configuration (i.e. the electron distribution in combination with the dust parameters, mainly) plays a crucial role in the existence and structural characteristics of dromions expected to occur in a given plasma environment.

\section{Application to Saturn's magnetosphere}

The Cassini mission has reported a large number of  in situ observations that have been used in many dusty plasma models. The powerful combination of 12 onboard instruments has provided unprecedented data on the planet's rings and their associated dynamical processes, composition, density, and size distribution.  Furthermore, these observations have established that electron statistics from Saturn's magnetosphere also obey kappa distribution(s) with small values of the kappa index ($\kappa \sim$ 2-6, usually) \citep{sch08}. \citet{pic15} analyzed the Cassini Radio and Plasma Wave Science  (RPWS) Wideband Receiver (WBR) data explicitly for the existence of bipolar structures  at 10$R_s$. 

The satellite observations and numerical parameters suggested by \citet{yar07} inspired us to explore the existence of  DIA dromions  in magnetized dusty plasma in the Saturnian magnetispheric environment. A complete parametric  analysis has been carried out in the previous section, establishing the fact that DIA  dromions may indeed occur in Saturn's magnetized environment. These theoretical predictions can be used in the  interpretation of future observations related to electrostatic structures not only in Saturn's environment, but also in other planetary environments.

\section{Conclusions}

We have analyzed a two-dimensional plasma fluid model for electrostatic (dust-ion-acoustic) excitations in a non-Maxwellian magnetized dusty plasma, taking into account the presence of a suprathermal electron population and of immobile dust in the background. The model was reduced to a Davey–Stewartson (DS) system, which was used as basis for further analysis of nonlinear envelope modes. We have shown that, under certain circumstances, multidimensional lump-shaped localized waveforms may occur, that can be modeled as dust-ion acoustic dromions, i.e. exact analytical solutions of the DS system, for specific sets of the relevant parameter values.

Our investigation has elucidated the existence conditions for dromions in a magnetized dusty plasma. The magnetic field plays a crucial role in providing the conditions for  the occurrence of dromions. The combined effects of various physical parameters such as the electron superthermality (spectral) index $\kappa$, the magnetic field strength, and the dust concentration on the dromion profile have been examined.
Our results are in agreement with (and in fact complement and generalize) recent studies by \citet{Singh2022} in the unmagnetized case and by  \citet{xue04} in the Maxwellian-electron unmagnetized case.

Focusing on observations in Saturn's magnetosphere, we have considered a set of parameter values adapted from Refs. \citep{goe89,yar07} to investigate whether dromion solutions may occur in the Saturnian magnetospheric environment.

Our results may be compared against existing experimental data in space, especially in Saturn's magnetosphere. \\

\vskip 1cm

\section*{Acknowledgements}

Authors KS and IK gratefully acknowledge financial support from Khalifa University of Science and Technology, Abu Dhabi UAE via the (internal funding) project FSU-2021-012/8474000352. Author IK gratefully acknowledges financial support from Khalifa University’s Space and Planetary Science Center under grant No. KU-SPSC-8474000336, in addition to support from KU via CIRA (Competitive Internal Research) CIRA-2021-064/8474000412 grant.

This work was initiated during a research visit by MMK to the Department of Sciences and Engineering, Sorbonne University Abu Dhabi, UAE where author IK was affiliated until 2019. Hospitality by the host (SUAD) is warmly acknowledged.
Authors MMK and IK acknowledge funding from the Abu Dhabi Department of Education and Knowledge (ADEK), currently ASPIRE UAE, via the AARE-2018 research grant ADEK/HE/157/18 in the period 2019-2022.


\section*{Data Availability}

The data underlying this article will be shared on reasonable request to the corresponding author.


\appendix

\section{Coefficients in the DS system (27)}

The (real) coefficients in the system of Eqs. (\ref{dssystem}) are given by:
\begin{multline} Q_1=\frac{1}{2\omega^2(c_1+k^2)}\left[(\omega^2-\Omega^2)\left(\omega\left(2c_2\left(\cad+\ccd\right)+3c_3\right)\right.\right. \\ \left.\left.-(\mu_n C_{24}^2+\mu_x(C_{21}^0+C_{21}^2))k_x\right)-\left(k_x^2 \omega \mu_x C_{24}^2+\frac{k_x^2 \Omega}{\omega}(2\mu_x C_{25}^2-\mu_y C_{24}^2)\right)\right]\nn\\   {\rm Here},\quad Q_1\lessgtr0, \, {\rm see\, Fig.} 1(g).\\
 Q_2= \frac{(\omega^2-\Omega^2)}{2\omega(c_1+k^2)} (2c_2 \omega \gamma_{\phi}-k_x \mu_x \gamma_n) \, , \nn\\  {\rm Here},\quad Q_2>0, {\rm see \, Fig.} 1(h).\\
Q_3=\left(2\frac{\mu_n\mu_x}{c_1v_{g,x}}-\frac{2c_2}{c_1}\right)=\cad    \, , \nn\\  {\rm Here}, \quad Q_3>0, \, {\rm see \, Fig.} 1(i).\\
P_1=\frac{1}{2}\frac{\partial^2 \omega}{\partial k_x^2}=\frac{1}{2\omega(c_1+k^2)}\left[(\omega^2-\Omega^2)(\rho_1-\omega-2k_x v_{g,x})\right.\\   \left.-\omega k_x\left(v_{g,x}\rho_1+\frac{\Omega v_{g,x}}{\omega}\rho_2\right)\right]   \, ,\nn\\ {\rm Here},\quad P_1\lessgtr0, \, {\rm see \,  Fig.} 1(c).\\
P_2 = \frac{1}{2}\frac{\partial^2 \omega}{\partial k_z^2}=\frac{c_1}{2\omega (c_1+k^2)^2}-\frac{\Omega^2}{2\omega^3 (c_1+k^2)^2}  \, , \nn\\ {\rm Here},\quad P_2>0, {\rm see \, Fig.} 1(d).\\
P_3 = -v_g \, , \qquad P_4 = \gamma_{\phi}\, , \nn\\ {\rm Here}, \quad P_3<0 \quad {\rm and} \quad  P_4>0 ; \, {\rm see \, \,  Fig.} 1(e,f). \, {\rm  Note \, that} \, v_{g} > 0. \\
\rho_1= \frac{\omega}{\omega^2-\Omega^2}\left(v_{g,x} \mu_x\left(1+\frac{\Omega^2}{\omega^2}\right)+1\right)\, , \nn\\
\rho_2= \frac{\Omega}{\omega}\rho_1+i\frac{\mu_y v_{g,x}}{\omega}\,    .\nn\\ \label{coefs} \end{multline}


\bsp	
\label{lastpage}
\end{document}